\documentclass[12pt]{iopart}

\usepackage{iopams}
\usepackage[dvipdfmx]{graphicx}
\usepackage[dvipdfmx]{color}
\usepackage{dcolumn}
\usepackage{bm}
\usepackage{color}
\usepackage{algorithm}
\usepackage{algorithmic}

\newcommand{\argmin}{\mathrm{argmin}}

\begin{document}

\title[Compressed sensing with quantum-classical hybrid approach]{L0 regularization-based compressed sensing with quantum-classical hybrid approach}

\author{Toru Aonishi$^1$, Kazushi Mimura$^{2,1}$, Masato Okada$^{3,1}$, Yoshihisa Yamamoto$^{4,5}$}
\address{$^1$School of Computing, Tokyo Institute of Technology, Yokohama, Kanagawa, Japan}
\address{$^2$Graduate School of Information Sciences, Hiroshima City University, Hiroshima, Japan}
\address{$^3$Graduate School of Frontier Sciences, The University of Tokyo, Kashiwa, Chiba, Japan}
\address{$^4$Physics and Informatics Laboratories, NTT Research Inc., Palo Alto, CA, USA}
\address{$^5$E. L. Ginzton Laboratory, Stanford University, Stanford, CA, USA}
\ead{aonishi@c.titech.ac.jp}

\vspace{10pt}
\begin{indented}
\item[]October 2021
\end{indented}

\begin{abstract}
L0-regularization-based compressed sensing (L0-RBCS) has the potential to outperform L1-regularization-based compressed sensing (L1-RBCS), but the optimization in L0-RBCS is difficult because it is a combinatorial optimization problem. To perform optimization in L0-RBCS, we propose a quantum-classical hybrid system consisting of a quantum machine and a classical digital processor. The coherent Ising machine (CIM) is a suitable quantum machine for this system because this optimization problem can only be solved with a densely connected network. To evaluate the performance of the CIM-classical hybrid system theoretically, a truncated Wigner stochastic differential equation (W-SDE) is introduced as a model for the network of degenerate optical parametric oscillators, and macroscopic equations are derived by applying statistical mechanics to the W-SDE. We show that the system performance in principle approaches the theoretical limit of compressed sensing and this hybrid system may exceed the estimation accuracy of L1-RBCS in actual situations, such as in magnetic resonance imaging data analysis.
\end{abstract}

\section{Introduction}
Quantum machines have attracted significant interest because of their potential to overcome the difficulty of solving large-scale combinatorial optimization problems. Many quantum machines, such as the quantum annealers (QA) of D-Wave systems \cite{RN843}, the quantum approximate optimization algorithm (QAOA) \cite{farhi2014quantum, PhysRevX.10.021067}, quantum bifurcation machines \cite{RN1589,RN1591,RN1594},    electromechanical resonators \cite{RN1596}   and coherent Ising machines (CIMs) \cite{RN846,RN892,RN899,RN813,RN1547,RN1548}, have been proposed in the past decade. Other examples include classical annealers, which have been implemented in nanomagnet arrays \cite{RN1598}, electronic oscillators \cite{RN1600}, silicon photonic weight banks \cite{RN1600}, complementary metal-oxide-semiconductor static random access memory circuits \cite{RN1602,7350099,8118124}, and field-programmable gate arrays (FPGAs) \cite{ChihiroYoshimura2017,RN1605}. Interest has been centered on implementing quantum machines and understanding their behavior, whereas there have been few practical applications \cite{RN1585,RN1571,RN1587}. To open the door to practical use of quantum machines, we show that they can be used for implementing compressed sensing (CS). Furthermore, we demonstrate, using non-equilibrium statistical mechanics \cite{RN848}, that the system performance in principle approaches the theoretical limit of CS.

L1-regularization-based CS (L1-RBCS) including the least absolute shrinkage and selection operator (LASSO) \cite{RN900} is a very efficient approach to solving various sparse signal reconstruction problems in exploration geophysics \cite{RN927,RN926,RN928,RN929}, magnetic resonance imaging (MRI) \cite{RN910,RN930,RN931,RN932}, black hole observation \cite{RN902}, and materials informatics \cite{RN933,RN934}. L1-RBCS is formulated as:
\begin{eqnarray}
x = \argmin_{x\in\mathbb{R}^N} \left(\frac{1}{2}\left\|y - A x \right\|^2_2 + \lambda \left\|x \right\|_1 \right), \label{eq.L1-regularization-based-CS}
\end{eqnarray}
where $x$ is an $N$-dimensional source signal, $y$ is an $M$-dimensional observation signal, $A$ is an $M$-by-$N$ observation matrix, and $\lambda$ is a regularization parameter.    Here, the ratio of the number of non-zero elements in the source signal $x$ to $N$ is defined as the sparseness $a$, and the ratio of $M$ to $N$ is defined as the compression ratio $\alpha$.   L1-RBCS can be formulated as a convex optimization problem, for which many efficient heuristic algorithms are available \cite{RN920,RN921,RN925,RN912,RN911,RN924}. 

On the other hand, L0-regularization-based CS (L0-RBCS) can be formulated with the L0 norm instead of the L1 norm \cite{RN909}:
\begin{eqnarray}
x = \argmin_{x\in\mathbb{R}^N} \left(\frac{1}{2}\left\|y - A x \right\|^2_2 + \lambda \left\|x \right\|_0 \right). \label{eq.L0-regularization-based-CS}
\end{eqnarray}
L0-RBCS, as defined in Eq. (\ref{eq.L0-regularization-based-CS}), can be equivalently reformulated as a two-fold optimization problem \cite{RN909,RN901}:
\begin{eqnarray}
(r,\sigma) = \argmin_{\sigma\in\{0,1\}^N} \argmin_{r\in\mathbb{R}^N} \left(\frac{1}{2}\left\|y - A \left(\sigma \circ r\right) \right\|^2_2 + \lambda \left\|\sigma \right\|_0 \right). \label{eq.L0-regularization-based-CS-support}
\end{eqnarray}
Here, the vector $r$ is the value of the $N$-dimensional source signal and each element $r_i$ in $r$ represents the real-number value of the $i$-th element in the source signal. The vector $\sigma$ is called a support vector, which represents the places of the non-zero elements in the $N$-dimensional source signal. The element $\sigma_i$ in $\sigma$ takes either $0$ or $1$ to indicate whether the $i$-th element in the source signal is zero or non-zero. The symbol $\circ$ denotes the Hadamard product. From the elementwise representation of Eq. (\ref{eq.L0-regularization-based-CS-support}), the Hamiltonian (or cost function) of L0-RBCS can be written as
\begin{eqnarray}
\mathcal{H} = \frac{1}{2} \sum_{i,j=1}^N \sum_{\mu=1}^M A_i^\mu A_j^\mu r_i r_j \sigma_i \sigma_j - \sum_{i=1}^N \sum_{\mu=1}^M y^\mu A_i^\mu r_i \sigma_i + \lambda \sum_{i=1}^N \sigma_i, \label{eq.L0-regularization-based-CS-Hamiltonian}
\end{eqnarray}
where $A_i^\mu$ is an element in an $M$-by-$N$ observation matrix $A$, and $y^\mu$ is an element in an $M$-dimensional observation signal. 

The minimization of $\mathcal{H}$ with respect to $r$ under the condition that $\sigma$ is fixed is the same as the problem of solving a system of simultaneous linear equations that gives the minimum point of the quadratic potential for $r$.    On the other hand, the minimization of $\mathcal{H}$ with respect to $\sigma$ under the condition that $r$ is fixed is the same as the problem of quadratic unconstrained binary optimization to find the ground state of a Hamiltonian of the two-state Potts model, where $-\sum_{\mu=1}^M A_i^\mu A_j^\mu r_i r_j$ can be considered to be the mutual interaction between $\sigma_i$ and $\sigma_j$. 

It has been suggested that L0-RBCS has the potential to outperform L1-RBCS, because L1 regularization imposes a shrinkage on variables over a threshold (soft-thresholding) but L0 regularization does not impose such a shrinkage (hard-thresholding) \cite{RN909}. However, the optimization of the support vector is a combinatorial optimization problem, which can be mapped into    a Potts model , as mentioned above. In this problem, there are a lot of meta-stable states because the effective interaction $-\sum_{\mu=1}^M A_i^\mu A_j^\mu r_i r_j$ induces frustration in    the Potts model   in the minimization of $\mathcal{H}$ with respect to $\sigma$ under the condition that $r$ is fixed. Thus, it is difficult to solve this kind of problem. Because of this difficulty, only a few approximation algorithms have been proposed and they only work under special conditions \cite{RN936,RN937,RN938}.    Note that the two-fold optimization problem for L0-RBCS is conceptually similar to Benders' decomposition \cite{Benders:1962vi}. However, Eq. (\ref{eq.L0-regularization-based-CS-Hamiltonian}) contains a quadratic programming part but does not contain a linear programing part; thus, our method is not strictly an example of Benders' decomposition. Furthermore, because in our method the non-linear part is a combinatorial optimization problem, our problem cannot not be made easier even if Benders' decomposition can be performed.  

\begin{figure}[t]
 \begin{center}
 \includegraphics[height=4cm]{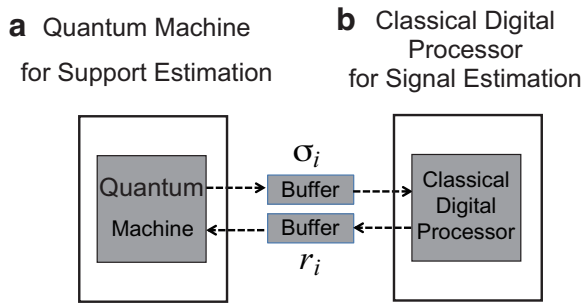}
 \end{center}
\caption{Quantum-classical hybrid system for L0-RBCS. To estimate the $N$-dimensional support vector $\sigma$ and $N$-dimensional signal vector $r$, this system solves a two-fold optimization problem by alternately performing two minimization processes; $\bf{a}$ the quantum machine optimizes $\sigma$ to minimize $\mathcal{H}$ with the given $r$, and $\bf{b}$ the classical digital processor optimizes $r$ to minimize $\mathcal{H}$ with the given $\sigma$.}
\label{fig:0}
\end{figure}

In this paper, to overcome the difficulty of optimizing the support vector $\sigma$, we focus on quantum machines. We propose a quantum-classical hybrid system composed of a quantum machine and a classical digital processor (CDP) (Fig. \ref{fig:0}). This system solves the two-fold optimization problem by alternately performing two minimization processes; (i) the quantum machine optimizes $\sigma$ to minimize $\mathcal{H}$ under the condition that $r$ is fixed, and (ii) the CDP optimizes $r$ to minimize $\mathcal{H}$ under the condition that $\sigma$ is fixed. If the quantum machine can find the ground state of $\mathcal{H}$ under the condition that $r$ is fixed, the quantum-classical hybrid system is expected to outperform L1-RBCS. 

Several quantum machines can potentially be used for optimizing $\sigma$, such as QA \cite{RN843}, QAOA \cite{farhi2014quantum, PhysRevX.10.021067}, CIM \cite{RN846,RN892,RN899,RN813,RN1547,RN1548}, and so on.    As defined in Eq. (\ref{eq.L0-regularization-based-CS-Hamiltonian}), the number of non-zero connections is $O(N^2)$; thus, it is necessary to form a densely connected network on a quantum machine in order to optimize $\sigma$.   A comparison of these candidates reveals that a measurement-feedback (MFB) CIM is one of most suitable machines for this purpose. In fact, an MFB-CIM can construct any densely connected network composed of degenerate optical parametric oscillators (OPOs) because it uses a time-division multiplexing scheme and MFB \cite{RN899,RN813}.    In contrast, QA and almost all other machines can only support local graphs, including chimera graphs, and thus, a densely-connected network for optimizing $\sigma$ has to be embedded in a fixed hardware local graph by using the minor-embedding scheme, which requires additional physical spins \cite{RN1576,RN1578}.   Furthermore, it was reported \cite{RN939} that an MFB-CIM experimentally outperformed QA on two problem sets, i.e., a fully connected Sherrington-Kirkpatrick model \cite{RN940} and dense graph MAX-CUT.    In contrast to QA having an exponential computation time proportional to $\exp\left(O\left(N\right)\right)$ , a CIM has an exponential computational time proportional to $\exp\left(O\left(\sqrt{N}\right)\right)$, where $N$ is the problem size \cite{RN939}.

\begin{figure}[t]
 \begin{center}
 \includegraphics[height=10cm]{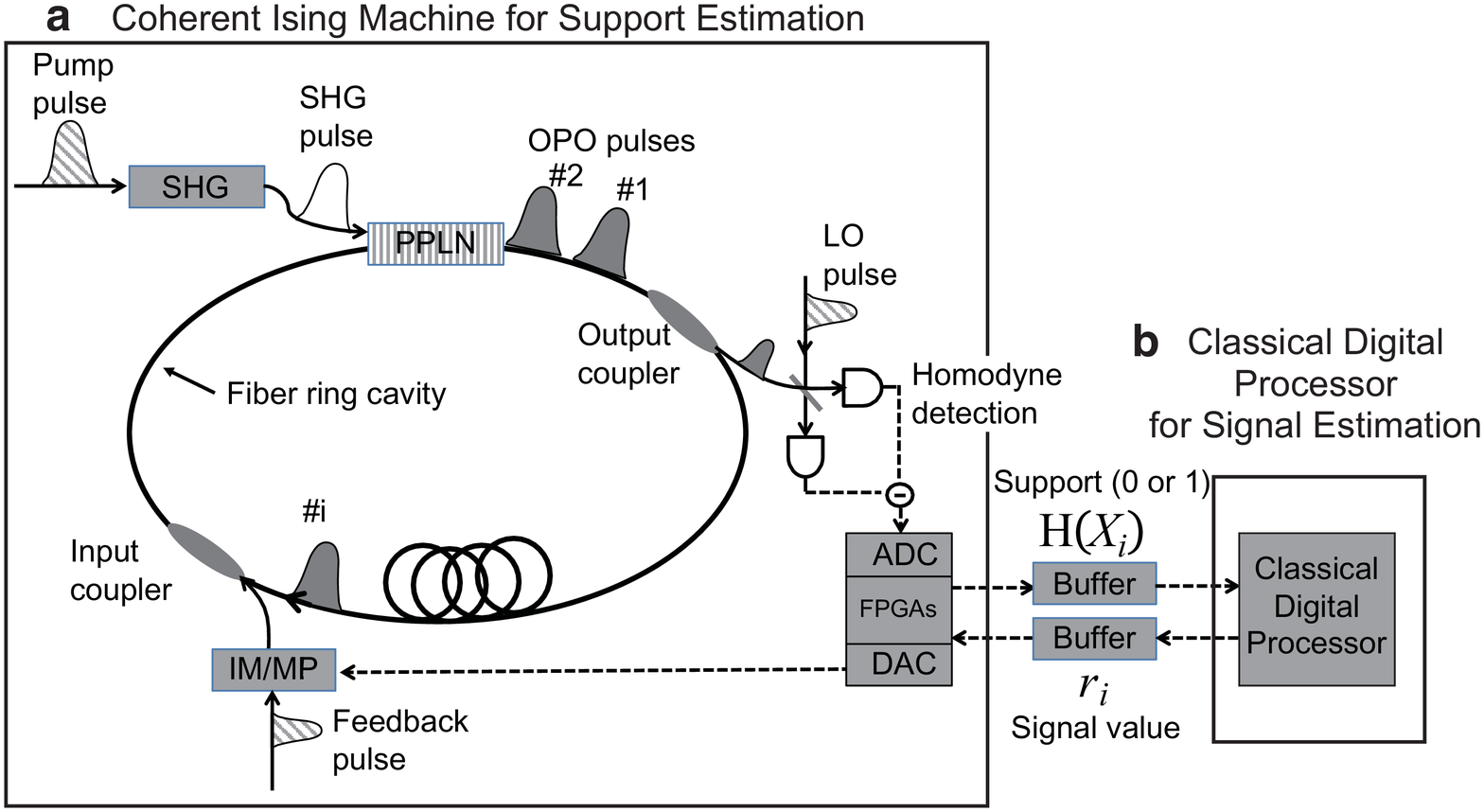}
 \end{center}
\caption{Quantum-classical hybrid system for L0-RBCS consisting of $\bf{a}$ coherent Ising machine (CIM) for support estimation and $\bf{b}$ classical digital processor (CDP) for signal estimation. This system performs the alternating minimization described in Algorithm \ref{alg1}. Pump pulses are injected into an optical parametric oscillator (OPO) formed in a fiber ring cavity through a second harmonic generation (SHG) crystal. A periodically poled lithium niobate (PPLN) waveguide device induces a phase-sensitive degenerate optical parametric amplification of the signal pulses, and each of the OPO pulses takes either the $0$-phase state (corresponding to the up-spin) or the $\pi$-phase state (corresponding to the down-spin) above the oscillation threshold. Part of each pulse is taken from the main cavity by the output coupler, and it is measured by optical homodyne detectors. A field programmable gate array (FPGA) calculates the feedback signal, which is then provided to the intensity modulator (IM) and phase modulator (PM) to produce the injection field described in Eq. (\ref{eq.optical-injection-field}) to each of the OPO pulses through the input coupler. $H(X_i)$ is a binarized value, either $0$ or $1$, of the in-phase amplitude of the $i$-th OPO pulse, which is the support estimate to be transferred to the CDP. The CDP solves the linear simultaneous equation (Eq. (\ref{eq.simultaneous-equations1})), and the solution $r_i$ is transferred to the CIM.}
\label{fig:1}
\end{figure}

Here, we evaluate the performance of a quantum-classical hybrid system composed of an MFB-CIM and CDP (Fig. \ref{fig:1}). We introduce a truncated Wigner stochastic differential equation (W-SDE) as a model for the network consisting of OPOs. Then, we develop a statistical mechanics method based on self-consistent signal-to-noise analysis (SCSNA) \cite{RN849,RN848,RN916} and derive a macroscopic equation (ME) for the whole system \cite{RN872,RN898,RN897}.   Several research groups have derived a critical condition for perfectly reconstructing $x$ in L$_p$ minimization-based CS (minimize $||x||_p$ s.t. $y=Ax$) when each entry of $A$ is an independently and identically distributed (i.i.d.) zero-mean Gaussian random number in the thermodynamic limit $N$, $M\to\infty$ with the compression rate $\alpha = M/N$ kept fixed \cite{doi:10.1073/pnas.0502258102,Kabashima_2009,RN905,RN901}. A threshold for the sparseness $a$ and the compression rate $\alpha$, called the weak threshold, determining whether or not the problem of L1-norm minimization has a solution with no error, was derived using techniques of combinatorial geometry \cite{doi:10.1073/pnas.0502258102}. On the other hand, the typical criticality of CS based on the general L$_p$ norm was explored, and thresholds for $p=0, 1, 2$, determining whether or not the problem of L$_p$-norm minimization has a solution with no error, were derived using statistical mechanics \cite{Kabashima_2009}. Note that the weak threshold derived with combinatorial geometry is perfectly consistent with the threshold for $p=1$ derived with statistical mechanics in the thermodynamic limit \cite{Kabashima_2009,RN905}. The role of the MEs derived here is mainly to show whether the theoretical performance limit of our model is comparable to the thresholds of L0/L1 minimization-based CS when the regularization parameter $\lambda$ is sufficiently small.   We show that the performance of the hybrid system approaches the theoretical limit of L0-minimization-based CS \cite{RN901} and the hybrid system may exceed the estimation accuracy of L1-RBCS in actual situations, such as MRI data analysis.

\section{Methods}
\subsection{Configuration of CIM-CDP hybrid system}
\label{sub2.1}
The CIM-CDP hybrid system (Fig. \ref{fig:1}) executes the L0-RBCS defined as Eqs. (\ref{eq.L0-regularization-based-CS-support}) and (\ref{eq.L0-regularization-based-CS-Hamiltonian}). This system optimizes by alternately performing the following two minimization processes. The CIM optimizes $\sigma$ to minimize $\mathcal{H}$ under the condition that $r$ is fixed and forwards $\sigma$ to the CDP. The CDP then optimizes $r$ to minimize $\mathcal{H}$ under the condition that $\sigma$ is fixed and then forwards $r$ to the CIM.

At a stationary point $r$ and $\sigma$ that satisfy $\frac{\partial \mathcal{H}}{\partial \sigma_i}=0$ and $\frac{\partial \mathcal{H}}{\partial r_i}=0$, the following equations hold   (see \ref{sec.maxwell})  :
\begin{eqnarray}
&\sigma_i& = H\left(r_i h_i - \lambda\right), \label{stationary_1} \\
&r_i& \sum_{\mu=1}^M \left(A_i^\mu\right)^2 = \sigma_i h_i, \ \ (i=1,\cdots,N) \label{stationary_2} \\
& & h_i = - \sum_{j=1(\neq i)}^N \sum_{\mu=1}^M A_i^\mu A_j^\mu \sigma_j r_j + \sum_{\mu=1}^M A_i^\mu y^\mu, \label{local_orgin}
\end{eqnarray}
where $h_i$ is the local field and $H(X)$ is the Heaviside step function taking $0$ for $X\leq 0$ or $+1$ for $X>0$.
$\lambda$ can be considered as the threshold.

In this paper, we assume that $\sum_{\mu=1}^M \left(A_i^\mu\right)^2 = 1$ is satisfied. This assumption does not lose any generality because it is possible to normalize the observation matrix $A$ to satisfy $\sum_{\mu=1}^M \left(A_i^\mu\right)^2 = 1$ for any case. Under this assumption, $r_i = \sigma_i h_i$ is satisfied in Eq. (\ref{stationary_2}), and according to the Maxwell rule \cite{RN917}, a stationary point of $\sigma_i = H\left(\sigma_i h^2_i - \lambda\right)$ obtained by substituting $r_i = \sigma_i h_i$ into Eq. (\ref{stationary_1}) can be determined as follows   (see \ref{sec.maxwell}) ,
\begin{eqnarray}
 \sigma_i &=& H\left(F_\chi(h_i) - \sqrt{2 \lambda}\right),\ \ (i=1,\cdots,N) \label{stationary_3} \\
 & &F_\chi(h) = \left\{ \begin{array}{l}
h \;\;\; ( \chi = + ) \\
|h| \;\; ( \chi = \pm )
\end{array}\right., \nonumber
\end{eqnarray}
  where the index $\chi$ of $F_\chi$ means whether the source signal is non-negative or signed and one of the functions, $F_{+}(h)$ or $F_{\pm}(h)$, is used depending on the source signal, as explained in \ref{sec.maxwell}.   $F_\chi (h)$ is the identity function if the source signal is non-negative, and $F_\chi (h)$ is the absolute value function if the source signal is signed. In the presence of noise, this conversion increases the threshold-to-noise ratio, which allows the low threshold to work as a sparse bias, as shown in the experiment below.

The CIM estimates the support vector $\sigma$, i.e. the places of the non-zero elements in the source signal. According to Eq. (\ref{stationary_3}), the optical field injected to the target ($i$-th) OPO pulse is set as
\begin{eqnarray}
f_i^{sig} &=& K \left(F_\chi(h_i^{CIM})-\eta \right),\ \ (i=1,\cdots,N) \label{eq.optical-injection-field} \\
 & &F_\chi(h) = \left\{ \begin{array}{l}
h \;\;\; ( \chi = + ) \\
|h| \;\; ( \chi = \pm )
\end{array}\right., \nonumber
\end{eqnarray}
where $h_i^{CIM}$ is the local field explained below, $K$ is the gain of the feedback circuit, and $\eta$ is the threshold. $\eta$ is related to $\lambda$ in Eqs. (\ref{eq.L0-regularization-based-CS-Hamiltonian}) and (\ref{stationary_1}) by $\eta=\sqrt{2\lambda}$, as shown in Eq. (\ref{stationary_3}). We use one of two functions, $F_{+}(h)$ or $F_{\pm}(h)$, depending on the source signal. $F_{+}(h)$ is the identity function: it is used as a non-negative source signal. $F_{\pm}(h)$ is the absolute value function: it is used as a signed source signal.

The local field for the support estimation in the CIM is set as 
\begin{eqnarray}
h_i^{CIM} = - \sum_{j=1(\neq i)}^M \sum_{\mu=1}^N A_i^\mu A_j^\mu r_j H(X_j) + \sum_{\mu=1}^M A_i^\mu y^\mu,\label{eq.localfield1}
\end{eqnarray}
where $r_j$ is a solution for the signal value given by the CDP, $X_j$ is the in-phase amplitude (generalized coordinate) of the $j$-th OPO pulse measured by a homodyne detector, and $H(X_j)$ is the binarized in-phase amplitude of the $j$-th OPO pulse through the Heaviside step function. The binarization of amplitude, which was proposed in the discrete simulated bifurcation \cite{doi:10.1126/sciadv.abe7953}, is necessary for improving the performance of the support vector estimation as described below. The first term of Eq. (\ref{eq.localfield1}) is the mutual interaction term, while the second term is the Zeeman term. During the support estimation on the CIM, all $r_j$ are fixed.
  
The support estimation in L0-RBCS is mathematically equivalent to the multi-user detector in code division multiple access (CDMA) \cite{RN881,RN897}. We reported that in the CDMA multi-user detector for the CIM, the system performance is not maximized unless the amplitude of the OPO pulse does not match the amplitude of the received sequence contained in the Zeeman term \cite{RN897}. Due to this equivalence to the CDMA multi-user detector, the mutual interaction term of Eq. (\ref{eq.localfield1}) can be considered to play a role in removing crosstalk noise evoked by the matched filter calculated in the Zeeman term. To remove the crosstalk noise completely, the amplitude of the OPO pulse $X$ needs to be the same as the amplitude of the elements of the source support vector, and thus, we binarize the value of $X$ in Eq. (\ref{eq.localfield1}) to take $1$ or $0$.  

The CDP estimates $r$, i.e. the values of the non-zero elements in the source signal. In accordance with the simultaneous equations (\ref{stationary_2}) satisfied by the stationary point that minimizes $\mathcal{H}$ with respect to $r$, the CDP solves the following simultaneous equations: 
\begin{eqnarray}
& &r_i\sum_{\mu=1}^M \left(A_i^\mu\right)^2 = H(X_i) h_i^{CDP},\ \ (i=1,\cdots,N) \label{eq.simultaneous-equations1}\\
& &h_i^{CDP} = -\sum_{j=1(\neq i)}^N \sum_{\mu=1}^M A_i^\mu A_j^\mu H(X_j) r_j + \sum_{\mu=1}^M A_i^\mu y^\mu.\label{eq.localfield2}
\end{eqnarray}
Here, $h_i^{CDP}$ in Eq. (\ref{eq.localfield2}) is the local field for the signal estimation in the CDP, and $H(X_j)$ is a solution for the support vector given by the CIM. During the signal estimation in the CDP, all $H(X_j)$ are fixed. The solution of the simultaneous equations (Eq. (\ref{eq.simultaneous-equations1})) is
\begin{eqnarray}
& &r = \left({\rm diag}[A^T A] + SA^TAS- {\rm diag}[SA^T AS]\right)^{-1} S A^T y, \nonumber \\
& &S = {\rm diag}\left(H(X_1), H(X_2), \cdots, H(X_N)\right). \nonumber
\end{eqnarray}

Algorithm \ref{alg1} is an outline of the alternating minimization process. In this algorithm, to make the basin of attraction wider, we heuristically introduce a linear threshold reduction whereby the threshold $\eta$ is linearly lowered from $\eta_{init}$ to $\eta_{end}$ as the alternating minimization proceeds.

During the support estimation on the CIM, all $r_j$ are fixed, while all $H(X_j)$ are updated in $h_i^{CIM}$. On the other hand, during the signal estimation on the CDP, all $H(X_j)$ are fixed, while all $r_j$ are updated in $h_i^{CDP}$. Therefore, $h_i^{CIM}$ becomes equal to $h_i^{CDP}$ when the whole system consisting of the CIM and CDP becomes steady.

\begin{algorithm}[b]
\caption{Alternating minimization of CIM-L0-RBCS}
\begin{algorithmic}[1]
\REQUIRE $M$-by-$N$ observation matrix: $A$, $M$-dimensional observation signal: $y$
 \ENSURE $N$-dimensional support vector: $\sigma$, $N$-dimensional signal vector: $r$
 \STATE Initialize $r=r_{init}$ and $\eta=\eta_{init}$
 \FOR{t=0 to 50}
 \STATE Minimize $\mathcal{H}$ with respect to $\sigma$ by using the CIM:\\
 \ \ \ \ \ $\sigma= {\rm CIM\_support\_estimation}(r,\eta)$\\
 \ $\#$ Initialize the c-amplitude as $c=0$, and numerically integrate the W-SDE while increasing the normalized pump rate from $0$ to $1.5$ for five times the photon's lifetime when $A_s^2=10^7$ or for two hundred times the photon's lifetime when $A_s^2=250$.
 \STATE Minimize $\mathcal{H}$ with respect to $r$ by using the CDP:\\
 \ \ \ \ \ $S={\rm diag}(\sigma)$\\
 \ \ \ \ \ $r=\left({\rm diag} [A^T A] +SA^T AS- {\rm diag}[SA^T AS]\right)^{-1} S A^T y$
 \STATE Decrement $\eta$: $\eta = {\rm max}(\eta_{init}(1-t/50),\eta_{end})$
 \ENDFOR
 \RETURN $\sigma$ and $r$
\end{algorithmic}
\label{alg1}
\end{algorithm}

\subsection{W-SDE for CIM}
Here, we introduce a CIM model consisting of $N$ OPO pulses coupled through the coherent feedback signal described in Eq. (\ref{eq.optical-injection-field}). By expanding the density operator of the whole OPO network with the Wigner function and applying Ito's rule to the resulting Fokker-Planck equation (see \ref{sec.W-SDE}), the following W-SDE can be derived.
\begin{eqnarray}
& &\frac{dc_i}{dt}=(-1+p-c_i^2-s_i^2)c_i+ \tilde{K}(F_\chi(h_i^{CIM})-\eta) + \frac{1}{A_s}\sqrt{c_i^2+s_i^2+1/2} W_{i,1},\nonumber\\
& &\frac{ds_i}{dt}=(-1-p-c_i^2-s_i^2 )s_i+ \frac{1}{A_s}\sqrt{c_i^2+s_i^2+1/2} W_{i,2},\ \ (i=1,\cdots,N)\label{eq.target-model}
\end{eqnarray}
where $c_i$ and $s_i$ are the in-phase and quadrature-phase normalized amplitudes of the $i$-th OPO pulse. $A_s$ is the saturation parameter which determines the nonlinear increase (abrupt jump) of the photon number at the OPO threshold.    The second term of the R.H.S. in the upper equation of Eq. (\ref{eq.target-model}) is the optical injection field corresponding to Eq. (\ref{eq.optical-injection-field}), which only has an in-phase component. The in-phase amplitude of the $i$-th OPO pulse, $X_i$, in Eq. (\ref{eq.localfield1}) is normalized as $c_i = X_i/A_s$, and $\tilde{K}$ is the normalized feedback gain corresponding to $K$.   $p$ is the normalized pump rate. $p=1$ corresponds to the oscillation threshold of a solitary OPO without mutual coupling. If $p$ is above the oscillation threshold ($p>1$), each of the OPO pulses is either in the $0$-phase state or $\pi$-phase state. The $0$-phase of an OPO pulse is assigned to an Ising-spin up-state, while the $\pi$-phase is assigned to the down-state. The last terms of the upper and lower equations express the vacuum fluctuations injected from external reservoirs and the pump fluctuations coupled to the OPO system via gain saturation. $W_{i,1}$ and $W_{i,2}$ are independent real Gaussian noise processes satisfying $\left<W_{i,k} (t)\right>=0$, $\left<W_{i,k} (t) W_{j,l} (t')\right> = \delta_{ij} \delta_{kl} \delta(t-t')$.

\subsection{Statistical mechanics}
\subsubsection{Precondition for applying statistical mechanics}
\label{precondition}
To solve the W-SDE (\ref{eq.target-model}) and the simultaneous equations (\ref{eq.simultaneous-equations1}) using statistical mechanics methods, we introduce the following observation model in which the values of all variables are randomly chosen,
\begin{eqnarray}
& & \left[\begin{array}{c}
 y^1 \\
 y^2 \\
 \vdots \\
 y^M
\end{array}
\right] = \frac{1}{\sqrt{M}}\left[
 \begin{array}{cccc}
 A_1^1 & A_2^1 & \ldots & A_N^1 \\
 A_1^2 & A_2^2 & \ldots & A_2^2 \\
 \vdots & \vdots & \ddots & \vdots \\
 A_1^M & A_2^M & \ldots & A_N^M
 \end{array}
 \right] \left[
 \begin{array}{c}
 \xi_1 x_1\\
 \xi_2 x_2\\
 \vdots \\
 \xi_N x_N
 \end{array}
 \right] + \left[
 \begin{array}{c}
 n^1 \\
 n^2 \\
 \vdots \\
 n^M
 \end{array}
 \right], \label{eq.observation model1}
\end{eqnarray}
where $[y^1,\cdots, y^M]^T$ is an $M$-dimensional observation signal, $[n^1,\cdots,n^M]^T$ is $M$-dimensional observation noise, $[x_1,\cdots, x_N]^T$ is an $N$-dimensional true source signal, and $[\xi_1,\cdots, \xi_N]^T$ is an $N$-dimensional true support vector. $[A_i^\mu]_{\mu=1,\cdots,M,i=1,\cdots,N}$ is the $M$-by-$N$ observation matrix, which is scaled by $1/\sqrt{M}$. Here, the compression rate $\alpha$ is defined as $\alpha=M/N$, as explained in the Introduction.    We will deal with the thermodynamic limit defined as the limit $N$, $M\to\infty$ with $\alpha=M/N$ kept fixed. 

Each element of $[A_i^\mu]_{\mu=1,\cdots,M,i=1,\cdots,N}$ is randomly generated and satisfies $\left<A_i^\mu\right>=0$ and $\left<A_i^\mu A_j^\nu\right>=\delta_{ij} \delta_{\mu\nu}$. Thus, $\frac{1}{M}\sum_{\mu=1}^M \left(A_i^\mu\right)^2 = 1$ is satisfied in the thermodynamic limit.

Each element of $[n^1,\cdots,n^M]^T$ is randomly generated, satisfying $\left<n^\mu\right>=0$ and $\left<n^\mu n^\nu\right>=\beta^2 \delta_{\mu\nu}$. $\beta^2$ is the variance of the observation noise. 

$aN$ elements in $[\xi_1,\cdots, \xi_N]^T$ are randomly selected and assigned $1$. Other elements are assigned $0$. Here, the sparseness $a$ is defined as the number of non-zero elements in the source signal, as explained in the Introduction. 

\begin{figure}[t]
\begin{center}
 \includegraphics[height=10cm]{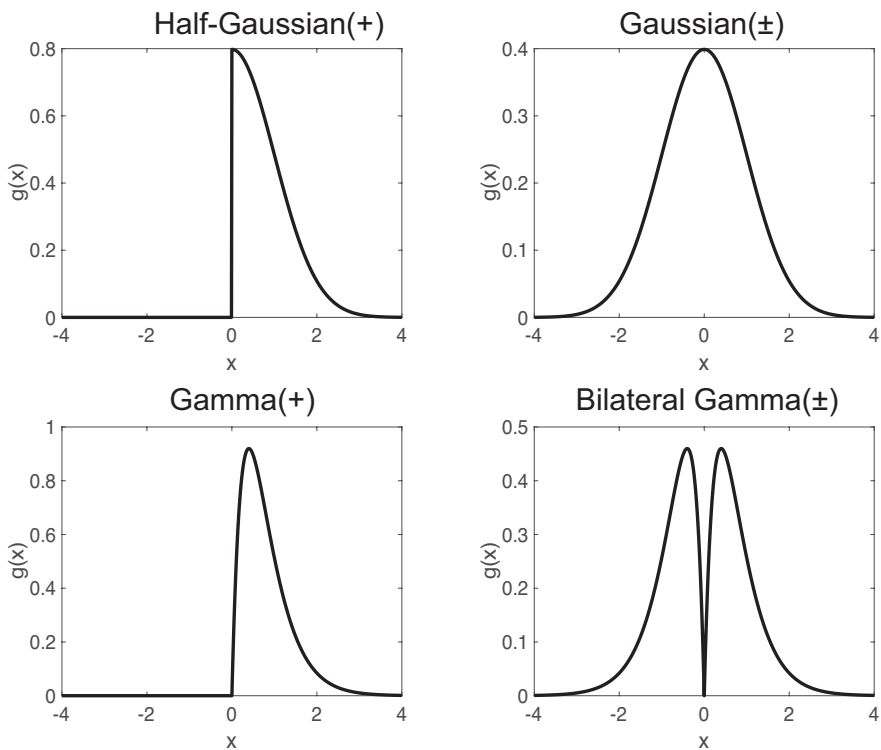}
 \end{center}
\caption{Four probability density functions used for generating the source signal in the numerical experiments. The half-Gaussian ($+$) and Gamma ($+$) are defined over a non-negative random variable. The Gaussian ($\pm$) and bilateral Gamma ($\pm$) are defined over a signed random variable. }
\label{fig:2}
\end{figure}

Each element of $[x_1,\cdots, x_N]^T$ is also an independent and identically distributed value generated from some probability distribution $g(x)$. To verify the system performance, we use the following probability density functions for generating the source signal: Gaussian($\pm$) $g(x)=e^{-x^2/{2\sigma^2}}/\sqrt{2\pi\sigma^2}$, half-Gaussian($+$) $g(x)=2H(x)e^{-x^2/{2\sigma^2}}/\sqrt{2\pi\sigma^2}$, Gamma($+$) $g(x)=x^{k-1} e^{-x/\theta}/(\Gamma(k)\theta^k)$, and bilateral Gamma($\pm$) $g(x)=|x|^{k-1} e^{-x/\theta}/(2\Gamma(k)\theta^k)$ (see Fig. \ref{fig:2}). To verify the invariance of our results relative to the type of probability distribution of the source signals, we used two different probability distributions in each of the non-negative and signed cases. Gaussian and bilateral Gamma distributions were used to generate the signed source signals. On the other hand, half-Gaussian and Gamma distributions were used to generate the non-negative source signals. The second moments of the half-Gaussian and Gaussian were set to $\left<x^2\right>_x=1$. The shape and scale parameters of the Gamma and bilateral Gamma were set to $k=2$ and $\theta=0.4$; thus, the second moment of both distributions was $\left<x^2\right>_x=0.96$. The figures in the main text show results for source signals generated from the half-Gaussian and Gaussian, while the supplementary figures show results for source signals from the Gamma and bilateral Gamma.

\subsubsection{Outline of derivation of MEs for the whole hybrid system}
\label{sec.outline}
Here, we summarize the derivation of the MEs by solving the W-SDE (\ref{eq.target-model}) and the simultaneous equations (\ref{eq.simultaneous-equations1}) under the precondition described in Section \ref{precondition}.    The procedure for deriving the MEs by applying SCSNA to the W-SDE for $N$ interacting OPO pulses is as follows \cite{RN849,RN848,RN916}.
\begin{enumerate}
\item	A formal transfer function from the local field to the unit output is introduced, and the local field is defined self-consistently through the formal transfer function. 
\item	Using the formal transfer function, the local field is decomposed into a pure local field and an Onsager reaction term (ORT). Then, the formal transfer function is redefined on the pure local field by renormalization of the ORT. Simultaneously, the macroscopic parameters, which are defined as the site average of the formal transfer function on the pure local field, are sought when decomposing the local field.
\item	By replacing the local field with the pure local field and the ORT, the W-SDE for $N$ interacting OPO pulses reduces to a system consisting of $N$ independent one-body OPO pulses. Then, the expectation of the formal transfer function on the pure local field is approximately derived from the one-body OPO pulse system.
\item	The site average of the formal transfer function on the pure local field, which defines the macroscopic parameter, is replaced with its expectation derived from the one-body OPO pulse system.
\item	Finally, the MEs are obtained.
\end{enumerate}
A detailed derivation of the MEs is provided in \ref{sub.ave} and \ref{sub.scsna}.
  
As described in Section \ref{sub2.1}, $h_i^{CIM}$ of Eq. (\ref{eq.localfield1}) becomes equal to $h_i^{CDP}$ of Eq. (\ref{eq.localfield2}) when the whole system consisting of the CIM and CDP becomes steady. If the pump power exceeds the oscillation threshold, the CIM reaches a steady state. Under this condition, the CIM and CDP share the same local field. Therefore, the CIM and CDP can be unified into a single mean field system in a steady state.
 
By substituting the observation model (\ref{eq.observation model1}), the shared local field can be rewritten as
\begin{eqnarray}
 h_i&=&-\frac{1}{M} \sum_{j=1(\neq i)}^M \sum_{\mu=1}^N A_i^\mu A_j^\mu r_j H(c_j) \nonumber \\
 & &+\frac{1}{M} \sum_{j=1}^M \sum_{\mu=1}^N A_i^\mu A_j^\mu \xi_j x_j +\frac{1}{\sqrt{M}} \sum_{\mu=1}^M A_i^\mu n^\mu.
\end{eqnarray}
where $X_j$ is replaced with $c_j$, which is the real part of the normalized complex Wigner amplitude $c_j+is_j$ explained in \ref{sec.W-SDE}. 

The W-SDE (\ref{eq.target-model}) of the $i$-th OPO implies that $H(c_i)$ is a stochastic variable depending on the local field $h_i$ and time $t$ in the steady state. Here, we introduce the following formal time-dependent stochastic transfer function from $h_i$ to $H(c_i)$ \cite{RN848,RN916}:
\begin{eqnarray}
H(c_i) = X(h_i,t). \nonumber
\end{eqnarray}
Substituting $X(h_i,t)$ into Eq. (\ref{eq.simultaneous-equations1}) and noting that $\frac{1}{M}\sum_{\mu=1}^M \left(A_i^\mu\right)^2 = 1$, we can write a formal time-dependent stochastic transfer function from $h_i$ to $r_i$ as
\begin{eqnarray}
r_i = G(h_i,t) = X(h_i,t) h_i. \nonumber
\end{eqnarray}
It is difficult to specify such a transfer function concretely, but it is possible to introduce one formally. As a premise that this transfer function holds, we assume that the microscopic memory effect can be neglected in the steady state \cite{PhysRevLett.88.024102}.

The local field can be defined self-consistently through the formal transfer function $G$, as follows:
\begin{eqnarray}
 h_i&=&-\frac{1}{M} \sum_{j=1(\neq i)}^M \sum_{\mu=1}^N A_i^\mu A_j^\mu G(h_j,t) \nonumber \\
 & &+ \frac{1}{M} \sum_{j=1}^M \sum_{\mu=1}^N A_i^\mu A_j^\mu \xi_j x_j + \frac{1}{\sqrt{M}} \sum_{\mu=1}^M A_i^\mu n^\mu,\label{eq.local-field3}
\end{eqnarray}
  
Given the formal transfer function, the local field $h_i$ can be separated into a pure local field and an ORT \cite{RN849,RN848,RN916} through manipulation of SCSNA in \ref{sub.scsna}.
\begin{eqnarray}
h_i = \tilde{h}_i + \Gamma H(c_i) r_i. \label{eq.local-field-ORT}
\end{eqnarray}
Then, $X(h_i,t)$ and $G(h_i,t)$ can be redefined with the pure local field $\tilde{h}_i$ by renormalizing the ORT as follows:
\begin{eqnarray}
H(c_i) = \tilde{X}(\tilde{h}_i,t), \ \ r_i= \tilde{G}(\tilde{h}_i,t) = \frac{1}{1-\Gamma} \tilde{h}_i \tilde{X}(\tilde{h}_i,t), \label{eq.transfer_for_PLF}
\end{eqnarray}
where $\tilde{X}$ and $\tilde{G}$ are formal time-dependent stochastic transfer functions from the pure local field $\tilde{h}_i$ to $H(c_i)$ and $r_i$, respectively. The pure local field $\tilde{h}_i$ and the coefficient of ORT $\Gamma$ can be self-consistently obtained as
\begin{eqnarray}
\tilde{h}_i=\frac{\alpha x_i \xi_i}{\alpha+aU} + \frac{\alpha}{\alpha+aU} \sqrt{\beta^2+\frac{a}{\alpha}(Q+\left<x^2\right>_x - 2R) } z_i, \ \ \Gamma=\frac{aU}{\alpha+aU},
\end{eqnarray}
where $z_i$ is Gaussian random noise obtained by separating the ORT from the cross-talk noise (see \ref{sub.scsna}), and $x_i$ and $\xi_i$ are the true source signal and true support described in Section \ref{precondition}. $\left<x^2\right>_x$ is the second moment of the source signal. $R$, $Q$ and $U$ are macroscopic parameters called the overlap, mean square magnetization, and susceptibility, respectively. $R$, $Q$ and $U$ are defined as follows.
\begin{eqnarray}
& &R=\frac{1}{aN} \sum_{j=1}^N x_j \xi_j \tilde{G}(\tilde{h}_i,t),\label{eq.overlap2}\\
& &Q=\frac{1}{aN} \sum_{j=1}^N \tilde{G}(\tilde{h}_i,t)^2.\label{eq.square_mag2} \\
& &U=\frac{1}{aN} \sum_{j=1}^N \frac{\partial \tilde{G}(\tilde{h}_j,t)}{\partial \tilde{h}_j} \frac{\partial \tilde{h}_j}{\partial h_j},\label{eq.susp2}
\end{eqnarray}
The overlap $R$ is the inner product between the source signal $x_j \xi_j$ and $\tilde{G}(\tilde{h}_i,t)$, the mean square magnetization $Q$ is the site average of the square of $\tilde{G}(\tilde{h}_i,t)$, and the susceptibility $U$ is the site average of the sensitivity of $\tilde{G}(\tilde{h}_i,t)$ to the bare local field $h_i$.
 
Through the SCSNA manipulation, the terms causing the correlation between OPO pulses are extracted by performing a first-order Taylor expansion and form the ORT and the scale coefficient $\alpha/(\alpha+aU)$ of the pure local field $\tilde{h}_i$ (see \ref{sub.scsna}). Therefore, the pure local field $\tilde{h}_i$ of the $i$-th OPO pulse is statistically independent of $\tilde{h}_j$ of the $j$-th OPO pulse when $i \neq j$. The ORT can be regarded as effective self-feedback via other OPO pulses.

It is difficult to specify the formal transfer function $\tilde{G}$ concretely, but it is possible to calculate the expectation as follows. By replacing the bare local field $h_i$ with the pure local field $\tilde{h}_i$ and the ORT $\Gamma H(c_i) r_i$, the W-SDE (\ref{eq.target-model2}) in \ref{sub.ave} can be regarded as describing $N$ independent one-body OPO pulses in the steady state, because the pure local fields $\tilde{h}_i$ are statistically independent of each other. Thus, the expectations $\left<\tilde{G}(\tilde{h},t)\right>_{\rm SDE}$, $\left<\tilde{G}(\tilde{h},t)^2\right>_{\rm SDE}$ and $\left<\partial \tilde{G}(\tilde{h},t)/\partial \tilde{h}\right>_{\rm SDE}$, which are the conditional expectations of $\tilde{G}(\tilde{h},t)$, $\tilde{G}(\tilde{h},t)^2$ and $\partial \tilde{G}(\tilde{h},t)/\partial \tilde{h}$ given the pure local field $\tilde{h}$, can be approximately derived from the one-body W-SDE (see \ref{sub.ave}). 

Because the pure local fields $\tilde{h}_i$ are statistically independent of each other, the site averages in $R$, $Q$ and $U$ can be replaced with the averages of $\left<\tilde{G}(\tilde{h},t)\right>_{\rm SDE}$, $\left<\tilde{G}(\tilde{h},t)^2\right>_{\rm SDE}$ and $\left<\partial \tilde{G}(\tilde{h},t)/\partial \tilde{h}\right>_{\rm SDE}$ with respect to the Gaussian random noise $z$ and the source signal $x \xi$ in $\tilde{h}$. 

Finally, the following MEs are obtained:
\begin{eqnarray}
&R&=\frac{1}{a} \int_{-\infty}^{+\infty} Dz \left<x\xi h_p \int_0^{+\infty} dc \int_{-\infty}^{+\infty} ds f(c,s|h_p) \right>_{x,\xi}, \label{eq.macro-eq1_1} \\
&Q&=\frac{1}{a} \int_{-\infty}^{+\infty} Dz \left<h_p^2 \int_0^{+\infty} dc \int_{-\infty}^{+\infty} ds f(c,s|h_p) \right>_{x,\xi},\label{eq.macro-eq1_2}\\
&U&\sqrt{\beta^2 + \frac{a}{\alpha}(Q+\left<x^2\right>_x-2R)}\nonumber \\
& &=\frac{1}{a}\int_{-\infty}^{+\infty} Dzz \left< h_p \int_0^{+\infty} dc \int_{-\infty}^{+\infty} ds f(c,s|h_p) \right>_{x,\xi}, \label{eq.macro-eq1_3}
\end{eqnarray}
where $\left<\cdot\right>_{x,\xi}$ denotes the average with respect to $x$ and $\xi$, and
\begin{eqnarray}
& &f(c,s|h_y) \propto \exp\left(\frac{2A_s^2\left(c\tilde{K}\left(F_\chi(h_y)-\eta\right)-V(c,s)\right)}{\Xi_c(z,x\xi)+\Xi_s (z,x\xi)+0.5}\right),\ (y=m,p) \nonumber\\
& &h_p=x\xi+\sqrt{\beta^2+\frac{a}{\alpha}(Q+\left<x^2\right>_x-2R)} z, \nonumber\\
& &h_m=\frac{1}{1+\frac{a}{\alpha} U} h_p, \nonumber\\
& &V(c,s)=\frac{1}{2} (1-p) c^2+\frac{1}{2} (1+p) s^2+ \frac{1}{2} c^2 s^2+\frac{1}{4}c^4+\frac{1}{4}s^4. \nonumber
\end{eqnarray}
$\Xi_c(z,x\xi)$ and $\Xi_s(z,x\xi)$ can be determined self-consistently from the following equations,
\begin{eqnarray}
& &\Xi_c (z,x\xi)=\int_{-\infty}^0 dc\int_{-\infty}^{+\infty} dsc^2 f(c,s|h_m)+\int_0^{+\infty}dc\int_{-\infty}^{+\infty} dsc^2 f(c,s|h_p),\nonumber\\
& &\Xi_s (z,x\xi)=\int_{-\infty}^0 dc\int_{-\infty}^{+\infty} dss^2 f(c,s|h_m)+\int_0^{+\infty}dc\int_{-\infty}^{+\infty} dss^2 f(c,s|h_p).\nonumber
\end{eqnarray}
$A_s$ is the saturation parameter, which diverges in the infinite limit of the amplitude of the injected pump field $\epsilon\to+\infty$ (see \ref{sec.W-SDE}). In the limit $A_s\to+\infty$, we obtain the following simplified MEs,
\begin{eqnarray}
&R&=\frac{1}{a} \int_{-\infty}^{+\infty} Dz \left<x\xi h_p \tilde{X}(h_p,h_m)\right>_{x,\xi}, \label{eq.macro-eq2_1}\\
&Q&=\frac{1}{a} \int_{-\infty}^{+\infty} Dz \left<h_p^2 \tilde{X}(h_p,h_m)\right>_{x,\xi},\label{eq.macro-eq2_2}\\
&U&\sqrt{\beta^2 + \frac{a}{\alpha}(Q+\left<x^2\right>_x-2R)}\nonumber\\
& &=\frac{1}{a} \int_{-\infty}^{+\infty} Dzz \left<h_p \tilde{X}(h_p,h_m)\right>_{x,\xi}.\label{eq.macro-eq2_3}
\end{eqnarray}
Here, $\tilde{X}(h_p,h_m)$ is an effective output function obtained from the Maxwell rule \cite{RN917}:
\begin{eqnarray}
\tilde{X}(h_p,h_m)=H(F_\chi(h_p)+F_\chi(h_m)-2\eta).\nonumber
\end{eqnarray}

\subsubsection{Perturbation expansion for ME in the limit $A_s\to\infty$ and $\eta\to +0$}
\label{sub.Perturbation}
By introducing a new macroscopic parameter $W$ defined by $W=Q-2R$, when there is no observation noise, i.e. $\beta=0$, we can rewrite the ME in the limit $A_s\to\infty$ as
\begin{eqnarray}
 &W&=\frac{a}{\alpha}\left|S+\left<x^2\right>_x\right|\frac{1}{a}\int_{-\infty}^{+\infty} Dz z^2 \left<\tilde{X}(h_p,h_m)\right>_{x,\xi} \nonumber \\
 & &-\frac{1}{a}\int_{-\infty}^{+\infty} Dz \left<x^2 \xi \tilde{X}(h_p,h_m)\right>_{x,\xi},\label{eq.macro-V}\\
& &\tilde{X}(h_p,h_m) = H\left(F_\chi(h_p)-2\eta/\left(1+1/F_\chi\left(1+\frac{a}{\alpha} U\right)\right)\right).\nonumber
\end{eqnarray}
Here, we put $2\eta/\left(1+1/F_\chi\left(1+\frac{a}{\alpha}U\right)\right)=\zeta^2$. In the limit $\eta\to +0$, i.e. $\zeta\to 0$, the ME (\ref{eq.macro-V}) has a solution $W=-\left<x^2\right>_x$ corresponding to perfect reconstruction.

We assume that the above ME has the following solution when $\zeta\ll 1$.
\begin{eqnarray}
W=-\left<x^2\right>_x + \zeta^2 w.\nonumber
\end{eqnarray}
Substituting this into the ME (\ref{eq.macro-V}) and expanding around $\zeta=0$, we obtain the following relation independent of the probability distribution of $x$, $g(x)$, if $g(0)$ and $g'(0)$ are finite.
\begin{eqnarray}
w = \frac{a}{\alpha}|w|.\nonumber
\end{eqnarray}
This equation suggests that the solution $w=0$, i.e. the perfect reconstruction solution, is stable when $a<\alpha$, neutral when $a=\alpha$, and unstable when $a>\alpha$.

Thus, when there is no observation noise, in the infinite limit of the amplitude of the injected pump field (i.e. $A_s^2\to\infty$) and in the infinitesimal limit of $\eta$, the critical point becomes $a=\alpha$ independent of $g(x)$.

\subsubsection{ME of LASSO}
Under the observation model (\ref{eq.observation model1}), the update rule of the LASSO is given by
\begin{eqnarray}
& &Y_i:=T_{\chi,\eta} (h_i),\ \ i=1,\cdots,N \nonumber \\
& &h_i=-\frac{1}{M} \sum_{j=1(\neq i)}^M \sum_{\mu=1}^N A_i^\mu A_j^\mu Y_j + \frac{1}{M} \sum_{j=1}^M \sum_{\mu=1}^N A_i^\mu A_j^\mu \xi_j x_j + \frac{1}{\sqrt{M}} \sum_{\mu=1}^M A_i^\mu n^\mu,\nonumber
\end{eqnarray}
where $T_{\chi,\eta} (h_j)$ is a soft-thresholding function with threshold $\eta$, defined as
\begin{eqnarray}
& &T_{+,\eta}(h)=\left\{ \begin{array}{ll}
h-\eta & (h\ge\eta) \\
0 & (h<\eta)
\end{array}\right.,\nonumber\\
& &T_{\pm,\eta}(h)=\left\{\begin{array}{ll}
h-\eta & (h\ge\eta) \\
0 & (-\eta<h<\eta) \\
h+\eta & (h\le-\eta)
\end{array}\right..\nonumber
\end{eqnarray}
We use two different functions $T_{+,\eta}(h)$ and $T_{\pm,\eta}(h)$ depending on the source signal. $T_{+,\eta}(h)$ is for non-negative source signals, and $T_{\pm,\eta}(h)$ is for signed source signals.

Following the same manipulation of the SCSNA, we obtain the following MEs,
\begin{eqnarray}
& &R=\frac{1}{a} \int_{-\infty}^{+\infty} Dz \left<x \xi \tilde{T}_{\chi,\eta}(\tilde{h})\right>_{x,\xi}, \label{eq.macro-eq-LASSO_1} \\
& &Q=\frac{1}{a} \int_{-\infty}^{+\infty} Dz \left<\tilde{T}_{\chi,\eta}(\tilde{h})^2\right>_{x,\xi},\label{eq.macro-eq-LASSO_2} \\
& &U\sqrt{\beta^2 + \frac{a}{\alpha}(Q+\left<x^2\right>_x-2R)}=\frac{1}{a}\int_{-\infty}^{+\infty} Dzz \left<\tilde{T}_{\chi,\eta}(\tilde{h})\right>_{x,\xi},\label{eq.macro-eq-LASSO_3}
\end{eqnarray}
where the pure local field of the LASSO becomes
\begin{eqnarray}
\tilde{h}_i=\frac{x \xi}{1+\frac{a}{\alpha}U} + \frac{\sqrt{\beta^2 + \frac{a}{\alpha}(Q+\left<x^2\right>_x-2R)}}{1+\frac{a}{\alpha}U} z.\nonumber
\end{eqnarray}
$\tilde{T}_{\chi,\eta}(\tilde{h})$ is an effective output function into which the ORT is renormalized:
\begin{eqnarray}
& &\tilde{T}_{+,\eta}(\tilde{h})=\left\{ \begin{array}{ll}
\left(1+\frac{a}{\alpha}U\right)(\tilde{h}-\eta) & (\tilde{h}\ge\eta) \\
0 & (\tilde{h}<\eta)
\end{array}\right.,\nonumber \\
& &\tilde{T}_{\pm,\eta}(\tilde{h})=\left\{\begin{array}{ll}
\left(1+\frac{a}{\alpha}U\right)(\tilde{h}-\eta) & (\tilde{h}\ge\eta) \\
0 & (-\eta<\tilde{h}<\eta) \\
\left(1+\frac{a}{\alpha}U\right)(\tilde{h}+\eta) & (\tilde{h}\le-\eta)
\end{array}\right..\nonumber
\end{eqnarray}
$\tilde{T}_{+,\eta}(\tilde{h})$ is for non-negative source signals, and $\tilde{T}_{\pm,\eta}(\tilde{h})$ is for signed source signals.

\subsection{Root-mean-square error}
\label{sub.Root-mean-square error}
The numerical experiments used the root-mean-square error (RMSE) as a measure of estimation accuracy. The RMSE of CIM L0-RBCS and LASSO is
\begin{eqnarray}
{\rm RMSE}=\sqrt{\frac{1}{N}\sum_{i=1}^N (r_i H(c_i) - x_i \xi_i)^2}=\sqrt{aQ-2aR+a\left<x^2\right>_x}, \nonumber
\end{eqnarray}
where $R$ and $Q$ are the overlap and mean square magnetization defined above, $a$ is sparseness, and $a\left<x^2\right>_x$ is the second moment of the source signal. The RMSE is zero if CIM L0-RBCS / LASSO perfectly reconstructs the source signal.

\section{Results}
\subsection{Evaluation of CIM L0-RBCS with statistical mechanics}
\begin{figure}[t]
 \begin{center}
 \includegraphics[height=13cm]{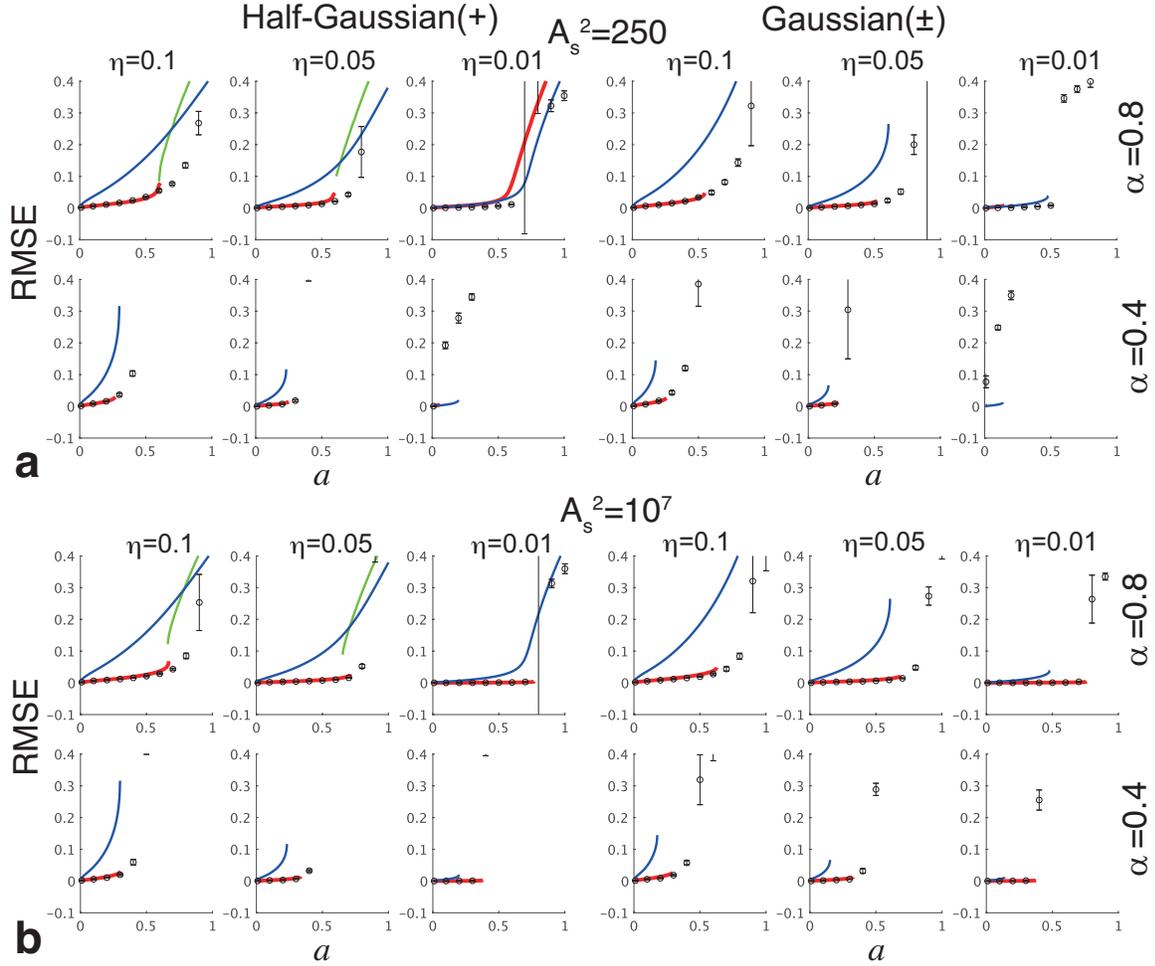}
 \end{center}
\caption{Comparison of solutions of the MEs with solutions of Algorithm \ref{alg1}: cases of no observation noise (i.e. $\beta=0$) and half-Gaussian ($+$) and Gaussian ($\pm$) source signals. The RMSEs of the solutions are plotted as a function of sparseness $a$ for various thresholds $\eta$ and compression rates $\alpha$. $\bf{a}$ Comparison of solutions of the MEs (\ref{eq.macro-eq1_1})(\ref{eq.macro-eq1_2})(\ref{eq.macro-eq1_3}) and those of Algorithm \ref{alg1} with $A_s^2=250$. $\bf{b}$ Comparison of solutions of the MEs (\ref{eq.macro-eq2_1})(\ref{eq.macro-eq2_2})(\ref{eq.macro-eq2_3}) for $A_s\rightarrow\infty$ and those of Algorithm \ref{alg1} with $A_s^2 =10^7$. In $\bf{a}$ and $\bf{b}$, the red and green lines respectively indicate RMSEs of the near-zero RMSE state and non-zero RMSE state in CIM L0-RBCS, which were obtained with the MEs (\ref{eq.macro-eq1_1})(\ref{eq.macro-eq1_2})(\ref{eq.macro-eq1_3}) with $A_s^2=250$ and the MEs (\ref{eq.macro-eq2_1})(\ref{eq.macro-eq2_2})(\ref{eq.macro-eq2_3}). The blue lines are the RMSEs of the near-zero RMSE state in LASSO, which were obtained with the MEs (\ref{eq.macro-eq-LASSO_1})(\ref{eq.macro-eq-LASSO_2})(\ref{eq.macro-eq-LASSO_3}) with the same threshold value of $\eta$ indicated above the graphs. The circles and error bars represent the mean values and standard deviations of ten trial solutions numerically obtained by Algorithm \ref{alg1}. To confirm the existence of solutions of Algorithm \ref{alg1} corresponding to the near-zero RMSE states indicated by the MEs, $r$ was initialized to the true signal value, i.e. $x \circ \xi$, and $\eta$ was kept constant by setting $\eta_{init}=\eta_{end}$ to the value of $\eta$ indicated above the graphs. For all graphs, $\tilde{K}=0.25$ and $N=2000$.}
\label{fig:3}
\end{figure}

\subsubsection{Typical solution of MEs, its accuracy, and comparison with LASSO when $\beta=0$}
\label{sec_me_sol}
 
First, several typical solutions of the MEs are shown for when there is no observation noise (i.e. $\beta=0$) and the source signals are from a half-Gaussian ($+$) or Gaussian ($\pm$). Moreover, to confirm the accuracy of the MEs, we compared the solutions to the MEs with those given by Algorithm \ref{alg1}. 

Figures \ref{fig:3}$\bf{a}$ and \ref{fig:3}$\bf{b}$ show the root-mean-square errors (RMSEs) (defined in Section \ref{sub.Root-mean-square error}) of the solutions to the MEs with $A_s^2=250$ (Eqs. (\ref{eq.macro-eq1_1})(\ref{eq.macro-eq1_2})(\ref{eq.macro-eq1_3})) and those in the limit $A_s^2\to\infty$ (Eqs. (\ref{eq.macro-eq2_1})(\ref{eq.macro-eq2_2})(\ref{eq.macro-eq2_3})) for various values of the threshold $\eta$ and compression rate $\alpha$ (red and green solid lines). The red line shows a solution whose RMSE increases monotonically from $0$ to some critical value as the sparseness $a$ increases from $0$ to some critical point $a_c$. On the other hand, the green line indicates a solution whose RMSE decreases monotonically from some finite value to some critical value as $a$ decreases from $1$ to some other critical point $a_c$. Here, the point at which the RMSE numerically calculated with the MEs discontinuously changes with increasing/decreasing $a$ is defined as the critical point $a_c$, and the RMSE at $a_c$ is defined as the critical value. In the following, the state indicated by the red line is called the near-zero-RMSE state, and the state indicated by the green line is called the non-zero-RMSE state. Regarding the results obtained by the MEs (\ref{eq.macro-eq2_1})(\ref{eq.macro-eq2_2})(\ref{eq.macro-eq2_3}), in the case of the half-Gaussian ($+$), two states, a non-zero-RMSE state (red solid line) and a near-zero one (green solid line), coexist, as in the CIM-implemented CDMA multiuser detector \cite{RN875,RN897}. On the other hand, in the case of the Gaussian ($\pm$), we numerically found only a near-zero RMSE state (red solid line). As shown by the red solid lines in Fig. \ref{fig:3}$\bf{b}$, in the limit $A_s^2\to\infty$, as $\eta$ was lowered to $0.01$, the RMSE of the near-zero-RMSE state decreased monotonically and the critical point $a_c$ from the near-zero-RMSE state grew monotonically. 

The circles and error bars in the figures indicate the mean and standard deviation of the RMSEs of ten trial solutions numerically obtained using Algorithm \ref{alg1} with $A_s^2=250$ and $A_s^2=10^7$. Note that $A_s^2=10^7$ is on the same order as $A_s^2$ in real experimental CIMs. To confirm if Algorithm \ref{alg1} has solutions corresponding to the near-zero-RMSE states obtained by the MEs, $r$ was initialized to the true signal value, i.e. $x\circ\xi$, in the alternating minimization process in Algorithm \ref{alg1}. In both the half-Gaussian case ($+$) and Gaussian case ($\pm$), the near-zero-RMSE states of the MEs (\ref{eq.macro-eq2_1})(\ref{eq.macro-eq2_2})(\ref{eq.macro-eq2_3}) (red solid lines in Fig. \ref{fig:3}$\bf{b}$) matched the numerical results of Algorithm \ref{alg1} with $A_s^2=10^7$ (circles with error bars in Fig. \ref{fig:3}$\bf{b}$), and the critical points given by the MEs (\ref{eq.macro-eq2_1})(\ref{eq.macro-eq2_2})(\ref{eq.macro-eq2_3}) coincided with those of Algorithm \ref{alg1}. On the other hand, the theoretical results obtained from the MEs (\ref{eq.macro-eq1_1})(\ref{eq.macro-eq1_2})(\ref{eq.macro-eq1_3}) with $A_s^2=250$ (red solid lines on the left of Fig. \ref{fig:3}$\bf{a}$) were in good agreement with the numerical results of Algorithm \ref{alg1} with $A_s^2=250$ (circles with error bars on the left of Fig. \ref{fig:3}$\bf{a}$) in the half-Gaussian case ($+$), whereas the critical points given by the MEs (\ref{eq.macro-eq1_1})(\ref{eq.macro-eq1_2})(\ref{eq.macro-eq1_3}) became lower than those of Algorithm \ref{alg1} when $\eta=0.01$ in the Gaussian case ($\pm$), as shown on the right of Fig. \ref{fig:3}$\bf{a}$.

Furthermore, to compare the abilities of CIM L0-RBCS and LASSO, we computed the RMSE profiles of LASSO using the MEs (\ref{eq.macro-eq-LASSO_1})(\ref{eq.macro-eq-LASSO_2})(\ref{eq.macro-eq-LASSO_3}) with the same threshold value as CIM L0-RBCS; these profiles are superimposed upon Fig. \ref{fig:3} (blue solid lines). The RMSEs of CIM L0-RBCS in the limit $A_s^2\to\infty$ (red solid lines in Fig. \ref{fig:3}$\bf{b}$) were lower than those of LASSO (blue solid lines) at the same compression rate $\alpha$ and sparseness $a$, and the critical points of CIM L0-RBCS were higher than those of LASSO. On the other hand, the RMSEs of CIM L0-RBCS with $A_s^2=250$ (red solid lines and circles with error bars in Fig. \ref{fig:3}$\bf{a}$) were lower than those of LASSO (blue solid lines) when $\eta=0.1$ and $0.05$, but the theoretical RMSEs of CIM L0-RBCS became higher than those of LASSO when $\eta=0.01$.

We numerically checked that qualitatively the same results were obtained even in the case of source signals from the Gamma ($+$) and the bilateral Gamma ($\pm$) (See Supplementary Fig. 1).

\begin{figure}[t]
\begin{center}
 \includegraphics[height=15cm]{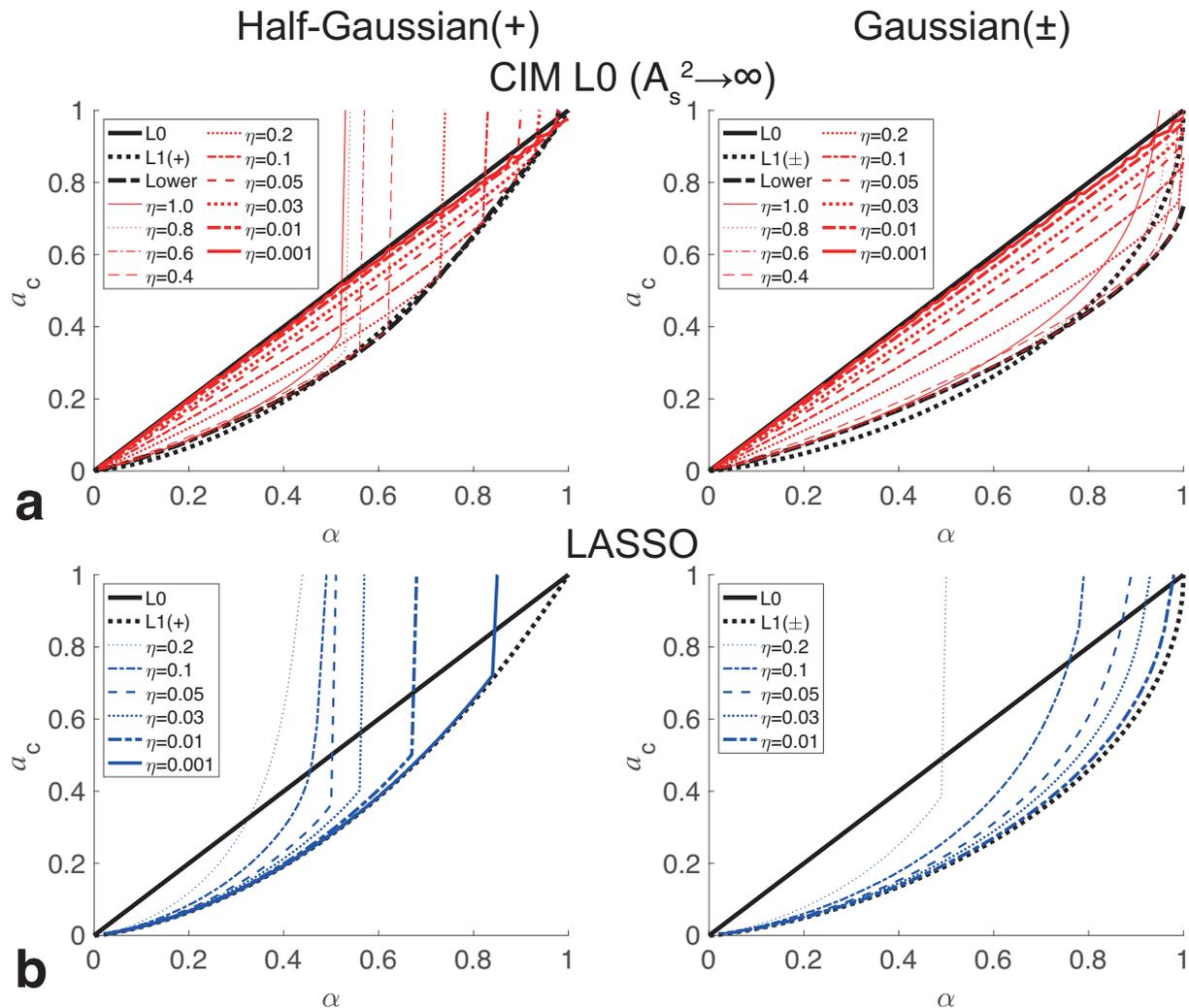}
 \end{center}
\caption{Phase diagrams of CIM L0-RBCS in the limit $A_s^2\to\infty$ and LASSO for various $\eta$: cases of no observation noise (i.e. $\beta=0$) and half-Gaussian ($+$) and Gaussian ($\pm$) source signals. $\bf{a}$ Phase diagrams of CIM L0-RBCS. $\bf{b}$ Phase diagrams of LASSO. In $\bf{a}$, the red lines show the critical point $a_c$ from the near-zero-RMSE state as a function of $\alpha$. The black dotted-dashed line in each plot indicates the lower bound of the critical points from the near-zero-RMSE state in CIM L0-RBCS. In $\bf{b}$, the blue lines show the critical point $a_c$ of LASSO as a function of $\alpha$.  The black solid line in each plot is a threshold determining whether or not the problem of L0-norm minimization CS has a solution with no error \cite{Kabashima_2009}, while the black dotted line is a threshold determining whether or not the problem of L1-norm minimization CS has a solution with no error for the non-negative case and signed case \cite{doi:10.1073/pnas.0502258102}.  }
\label{fig:4}
\end{figure}

\subsubsection{Phase diagrams of CIM L0-RBCS and LASSO when $\beta=0$}
\label{sec.phase_dia}
 
We drew phase diagrams of CIM L0-RBCS for various values of $\eta$ when there was no observation noise (i.e. $\beta=0$). The red lines in Figure \ref{fig:4}$\bf{a}$ and Supplementary Fig. 2 show the critical points from the near-zero-RMSE state (whose definition is given in Section \ref{sec_me_sol}) in the half-Gaussian case ($+$) and Gaussian case ($\pm$). The critical points in Fig. \ref{fig:4}$\bf{a}$ are for the limit $A_s^2\to\infty$, while the ones in Supplementary Fig. 2 are for $A_s^2=250$. To compare the properties of CIM L0-RBCS with those of LASSO, Fig. \ref{fig:4}$\bf{b}$ shows the phase diagrams of LASSO; the blue lines are the critical points from the near-zero-RMSE state for various $\eta$. If there is no discontinuous change in RMSE in $0\leq a\leq 1$, the critical point is not drawn on the phase diagrams. 

Several research groups have derived thresholds for determining whether or not the problem of L$_p$ minimization-based CS (minimize $||x||_p$ s.t. $y=Ax$) has a solution with no error. In particular, thresholds were derived for when each entry of $A$ is an i.i.d. zero-mean Gaussian random number in the thermodynamic limit $N$, $M\to\infty$ with $\alpha=M/N$ kept fixed \cite{doi:10.1073/pnas.0502258102,Kabashima_2009,RN905,RN901}, which is the same condition as the precondition in this paper. To confirm whether the theoretical performance limit of CIM L0-RBCS is comparable to the thresholds of L0/L1 minimization-based CS in the thermodynamic limit, below we compare the critical points of CIM L0-RBCS with the thresholds of L0/L1 minimization-based CS.

The threshold of L0 minimization-based CS is given by \cite{Kabashima_2009,RN901}
\begin{eqnarray}
a_{th}=\alpha. \nonumber 
\end{eqnarray}
The threshold $a_{th}$ of L0 minimization-based CS as a function of the compression rate $\alpha$ is shown by the black solid lines in Fig. \ref{fig:4}. If $\alpha>a$, a no-error solution is stable in L0 minimization-based CS in the thermodynamic limit. Note that the existence of a no-error solution was proved, but the performance of a specific algorithm for finding the solution was not shown in \cite{Kabashima_2009,RN901}. As demonstrated in Fig. \ref{fig:4}$\bf{a}$, in the limit $A_s^2\to\infty$, the critical points of CIM L0-RBCS become asymptotic to the black solid line as $\eta$ decreases and the RMSEs of CIM L0-RBCS are asymptotic to zero (the red lines in Fig. \ref{fig:3}$\bf{b}$). Thus, as $\eta$ decreases, the typical criticality of CIM L0-RBCS is asymptotic to that of L0 minimization-based CS. This result shows that, as $\eta$ decreases, the theoretical performance limit of CIM L0-RBCS in principle approaches the threshold of L0-minimization-based CS.

The threshold of L1 minimization-based CS, i.e., the weak threshold, is given by \cite{doi:10.1073/pnas.0502258102,RN905,Kabashima_2009},
\begin{eqnarray}
a_{th} = \alpha \max_{z \geq 0} \left\{\frac{1-\left(\kappa_\chi/\alpha\right)\left(1+z^2\Phi(-z)-z\phi(z)\right)}{1+z^2-\kappa_\chi\left(1+z^2\Phi(-z)-z\phi(z)\right)}\right\} \leq \alpha, \ \ (0\leq \alpha \leq 1), \nonumber
\end{eqnarray}
where $\kappa_\chi = 1, 2$, for the non-negative model ($\chi=+$) and the signed model ($\chi=\pm$), respectively. $\phi(z)$ is the standard Gaussian distribution and $\Phi(z)$ is the cumulative Gaussian distribution. The threshold $a_{th}$ of L1-minimization-based CS as a function of the compression rate $\alpha$ is shown by the black dotted lines in Fig. \ref{fig:4}. If $a_{th} (\alpha) > a$, a no-error solution is stable in L1 minimization-based CS. As $\eta$ decreases, the critical points of LASSO for the half-Gaussian ($+$) and Gaussian ($\pm$) become asymptotic to the two black dotted lines, and the RMSEs of LASSO become asymptotic to zero (the blue lines in Fig. \ref{fig:3}). Thus, as $\eta$ decreases, the typical criticality of LASSO become asymptotic to that of L1-minimization-based CS. On the other hand, the critical point of CIM L0-RBCS goes beyond the threshold of L1 minimization-based CS as $\eta$ decreases.
 
CIM L0-RBCS and LASSO have these asymptotic properties even when the source signals are from the Gamma ($+$) and bilateral Gamma ($\pm$) (see Supplementary Fig. 3$\bf{b}$). Note that we have theoretically proved that the asymptotic property of CIM L0-RBCS is invariant to differences in the probability distributions of the source signal by applying a perturbation expansion to the MEs (\ref{eq.macro-eq2_1})(\ref{eq.macro-eq2_2})(\ref{eq.macro-eq2_3}) in the limit $\eta\to +0$ (see Section \ref{sub.Perturbation}). Thus, we have confirmed this theoretical result numerically.

On the other hand, when $A_s^2=250$, the critical points of CIM L0-RBCS are not asymptotic to the black solid line $a=\alpha$, as shown in Supplementary Figs. 2 and 3$\bf{a}$. Around $\eta=0.1$, the critical point is closest to $a=\alpha$.

The black dotted-dashed lines in Fig. \ref{fig:4}$\bf{a}$ shows the lower bounds of the critical points of CIM L0-RBCS in the limit $A_s^2\to\infty$. The lower bound lines are above the threshold (black dotted line) of L1-minimization-based CS when the compression rate $\alpha$ is lower than around 0.5 for the half-Gaussian ($+$) and 0.7 for the Gaussian ($\pm$). The lower boundary property in Fig. \ref{fig:4}$\bf{a}$ is satisfied even in the case of source signals from the Gamma ($+$) and bilateral Gamma ($\pm$) (see Supplementary Fig. 3$\bf{b}$). On the other hand, there are no such lower bounds when $A_s^2=250$ (Supplementary Figs. 2 and 3$\bf{a}$).

\begin{figure}[t]
\begin{center}
 \includegraphics[height=15cm]{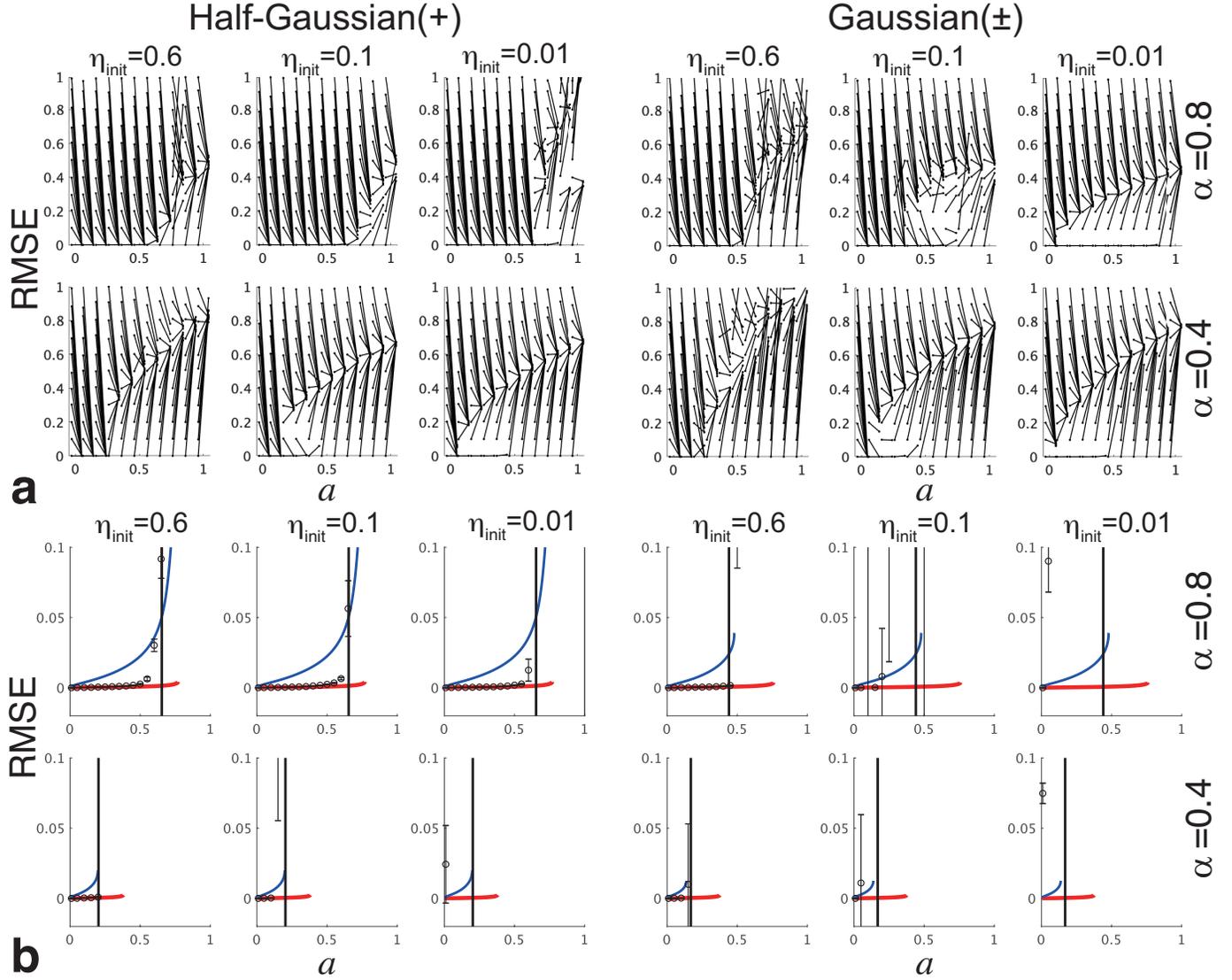}
 \end{center}
\caption{Basin of attraction of CIM L0-RBCS depending on the initial threshold $\eta_{init}$: cases of no observation noise (i.e. $\beta=0$) and half-Gaussian ($+$) and Gaussian ($\pm$) source signals.    $\bf{a}$ Size of the basin of attraction of Algorithm \ref{alg1} for various $\eta_{init}$ under fixed $\eta_{end}=0.01$. Pairs of points connected by a line indicate RMSEs of the initial and final states $\sigma\circ r$ of Algorithm \ref{alg1} for $a=0$ to $1$ in $0.1$ increments. $A_s^2=10^7$. $\alpha=0.4$ and $0.8$.  $\bf{b}$ Final states of Algorithm \ref{alg1} when starting from an initial state $r=0$ for various $\eta_{init}$. The circles and error bars represent the mean values and standard deviations of twenty trial solutions numerically obtained by Algorithm \ref{alg1} with $\eta_{end}=0.01$ and $A_s^2=10^7$. The red lines show the solutions of the MEs (\ref{eq.macro-eq2_1})(\ref{eq.macro-eq2_2})(\ref{eq.macro-eq2_3}) with near-zero RMSE when $\eta=0.01$ and $A_s^2\to\infty$, while the blue lines indicate the RMSEs of LASSO when $\eta=0.01$. The black lines are the lower bounds of the critical points of the CIM L0-RBCS. $\tilde{K}=0.25$ and $N=4000$.}
\label{fig:5}
\end{figure}

\subsubsection{Basin of attraction when $\beta=0$}
To check the practicality of CIM L0-RBCS, we verified the basin of attraction of Algorithm \ref{alg1}. To make the basin wider, we heuristically introduced a linear threshold attenuation wherein the threshold $\eta$ was linearly lowered from $\eta_{init}$ to $\eta_{end}$ as the minimization process was alternated (see Algorithm \ref{alg1}). First, we carried out numerical experiments to verify the size of the basin of attraction for various initial values $\eta_{init}$ for fixed $\eta_{end}=0.01$ in the case of no observation noise (i.e. $\beta=0$). As shown in Fig. \ref{fig:5}$\bf{a}$, the basin of attraction tended to be widened by selecting a higher initial threshold $\eta_{init}$ than $\eta_{end}$. As the compression rate $\alpha$ decreased, this tendency became more marked, especially in the Gaussian case ($\pm$). 

Next, we sought to confirm how well Algorithm \ref{alg1} converged on the near-zero RMSE state given by the MEs (\ref{eq.macro-eq2_1})(\ref{eq.macro-eq2_2})(\ref{eq.macro-eq2_3}) when starting from an initial state $r=0$ for various $\eta_{init}$ (Fig. \ref{fig:5}$\bf{b}$). As demonstrated in Fig. \ref{fig:5}$\bf{b}$, when the sparseness $a$ was lower than the lower bound of the critical points (the black dotted-dashed line in Fig. \ref{fig:4}$\bf{a}$), Algorithm \ref{alg1} with $\eta_{init}=0.6$ converged to the solutions (red lines) of the MEs (\ref{eq.macro-eq2_1})(\ref{eq.macro-eq2_2})(\ref{eq.macro-eq2_3}), whereas it failed to converge to the solutions for other values of $\eta_{init}$. Compared with the RMSE profiles of LASSO in Fig. \ref{fig:5}$\bf{b}$, Algorithm \ref{alg1} exceeded LASSO's estimation accuracy under almost all of the conditions in which LASSO had a small error.

The properties for the source signals taken from the Gamma ($+$) and bilateral Gamma ($\pm$) distributions (see Supplementary Fig. 4) are similar to those in Fig. \ref{fig:5}.

\begin{figure}[t]
 \begin{center}
 \includegraphics[height=14cm]{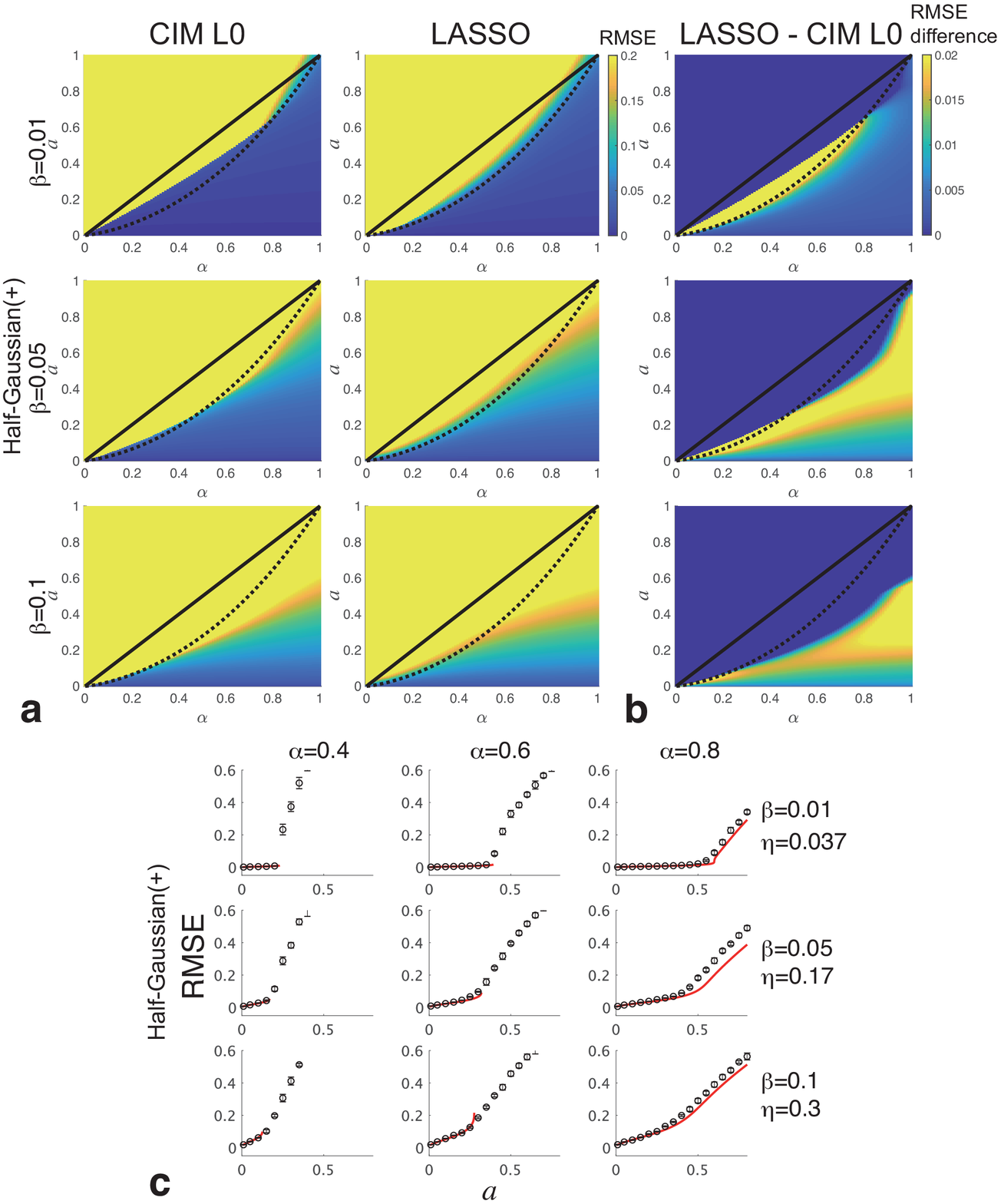}
 \end{center}
\caption{RMSEs under the optimal threshold when there is observation noise: case of half-Gaussian ($+$) source signals. The standard deviation of the observation noise was set to $\beta=0.01$, $0.05$ and $0.1$. $\bf{a}$ Comparison of RMSEs of CIM L0-RBCS and those of LASSO under the optimal threshold for each method. The color scale indicates the minimum RMSE under the optimal threshold at each point $(a,\alpha)$, which was obtained by a grid search for the set of solutions to the MEs (\ref{eq.macro-eq2_1})(\ref{eq.macro-eq2_2})(\ref{eq.macro-eq2_3}) and the MEs (\ref{eq.macro-eq-LASSO_1})(\ref{eq.macro-eq-LASSO_2})(\ref{eq.macro-eq-LASSO_3}) in the range $0.002\leq \eta \leq 0.5$ at each point $(a,\alpha)$. $\bf{b}$ Difference in minimum RMSE between LASSO and the CIM L0-RBCS under the optimal threshold for each method. The color scale indicates the minimum RMSE of CIM L0-RBCS subtracted from that of LASSO at each point $(a,\alpha)$. $\bf{c}$ Comparison of solutions of the MEs (\ref{eq.macro-eq2_1})(\ref{eq.macro-eq2_2})(\ref{eq.macro-eq2_3}) and those of Algorithm \ref{alg1} with $A_s^2=10^7$. The red solid lines show the near-zero RMSE solutions to the MEs (\ref{eq.macro-eq2_1})(\ref{eq.macro-eq2_2})(\ref{eq.macro-eq2_3}). The circles and error bars represent the mean values and standard deviations of ten trial solutions numerically obtained by Algorithm \ref{alg1} when starting from the initial state $r=0$. The value of $\eta$ indicated on the right side of the graphs in $\bf{c}$ is the optimal threshold at $\alpha=0.5$, which was set as $\eta_{end}$. For all the graphs in $\bf{c}$, $\eta_{init}=0.6$, $\tilde{K}=0.25$ and $N=4000$.}
\label{fig:6}
\end{figure}

\begin{figure}[t]
 \begin{center}
 \includegraphics[height=14cm]{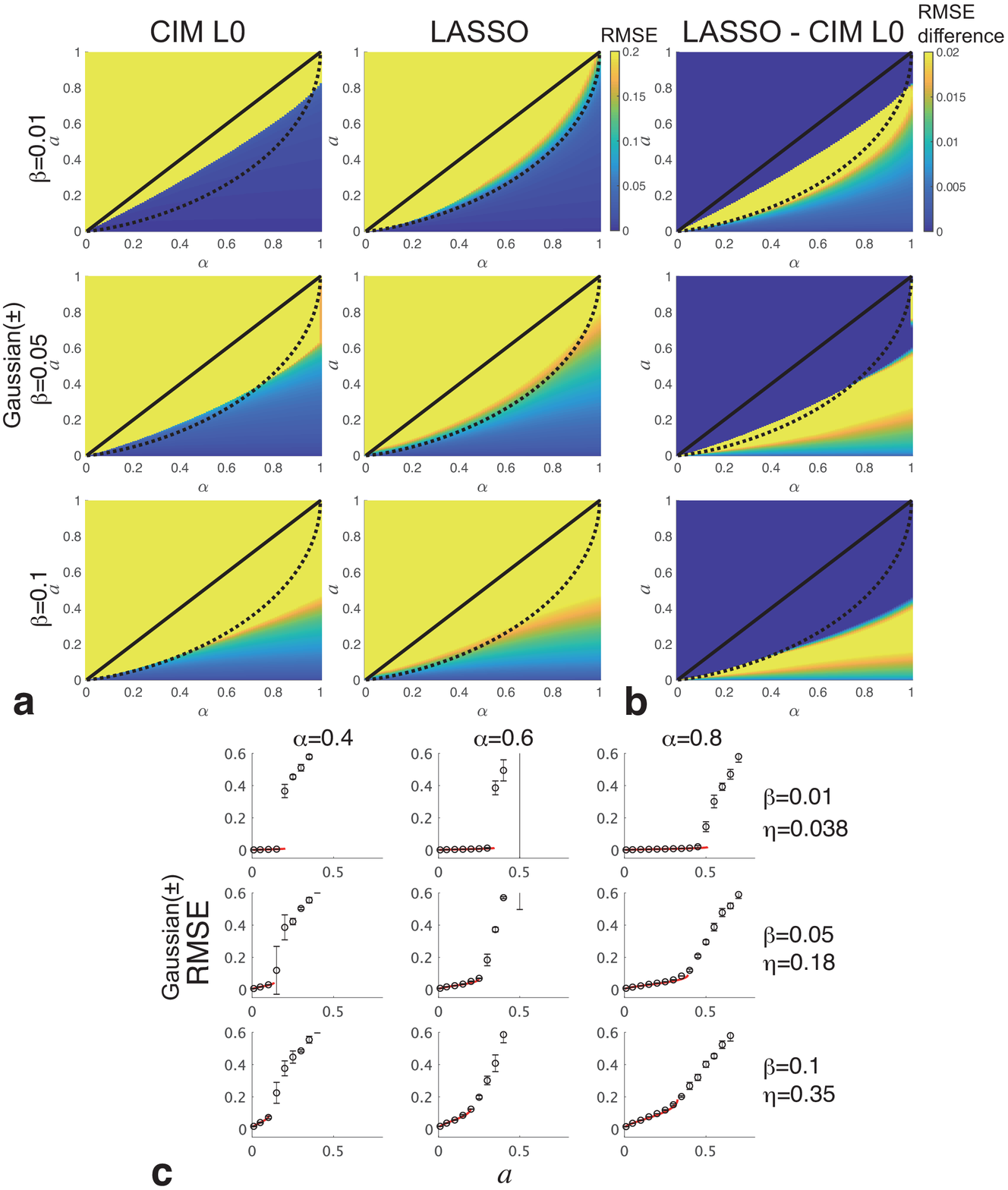}
 \end{center}
\caption{RMSEs under the optimal threshold when there is observation noise: case of Gaussian ($\pm$) source signals. The standard deviation of the observation noise was set to $\beta=0.01$, $0.05$, and $0.1$. The methods and conditions for obtaining these graphs are the same as in Fig. \ref{fig:6} except for the probability distribution of the source signals. $\bf{a}$ Comparison of RMSEs of CIM L0-RBCS and those of LASSO under the optimal threshold for each method. The color scale indicates the minimum RMSE under the optimal threshold at each point $(a,\alpha)$. $\bf{b}$ Difference in minimum RMSE between LASSO and the CIM L0-RBCS under the optimal threshold for each method. The color scale indicates the minimum RMSE of CIM L0-RBCS subtracted from that of LASSO at each point $(a,\alpha)$. $\bf{c}$ Comparison of solutions of the MEs (\ref{eq.macro-eq2_1})(\ref{eq.macro-eq2_2})(\ref{eq.macro-eq2_3}) and those of Algorithm \ref{alg1} with $A_s^2=10^7$ when starting from the initial state $r=0$.}
\label{fig:7}
\end{figure}

\subsubsection{Performance of CIM L0-RBCS and LASSO when $\beta\neq 0$}
Moreover, to check the practicality of CIM L0-RBCS, we verified its accuracy and convergence in the presence of observation noise (i.e. $\beta\neq 0$). We searched for the optimal threshold values that would give the minimum RMSEs of CIM L0-RBCS and those of LASSO (Figs. \ref{fig:6}$\bf{a}$ and \ref{fig:7}$\bf{a}$) and computed the difference between their minimum RMSEs (Figs. \ref{fig:6}$\bf{b}$ and \ref{fig:7}$\bf{b}$) under the optimal threshold for each method when $\beta=0.01$, $0.05$, and $0.1$. The minimum RMSE was obtained by conducting a grid search on the set of solutions to the MEs (\ref{eq.macro-eq2_1})(\ref{eq.macro-eq2_2})(\ref{eq.macro-eq2_3}) and the MEs (\ref{eq.macro-eq-LASSO_1})(\ref{eq.macro-eq-LASSO_2})(\ref{eq.macro-eq-LASSO_3}) in the range $0.002\leq \eta \leq 0.5$ at each point $(a,\alpha)$. These figures show cases of the half-Gaussian ($+$) and Gaussian ($\pm$) source signals. As indicated in Figs. \ref{fig:6}$\bf{a}$ and \ref{fig:7}$\bf{a}$, as $\beta$ decreases, the critical points from the-near-zero RMSE state in CIM L0-RBCS under the optimal threshold approaches the critical line (black solid line) of L0-minimization-based CS, and the RMSEs of CIM L0-RBCS under the optimal threshold decreases. As shown in Figs. \ref{fig:6}$\bf{b}$ and \ref{fig:7}$\bf{b}$, the RMSEs of LASSO are higher than those of CIM L0-RBCS under almost all of the conditions in which LASSO has an error less than $0.2$; thus, CIM L0-RBCS exceeds LASSO's estimation accuracy under the optimal threshold for each method.

Next, for the case of observation noise, we determined whether the output of Algorithm \ref{alg1} with $A_s^2=10^7$ converged on solutions to the MEs (\ref{eq.macro-eq2_1})(\ref{eq.macro-eq2_2})(\ref{eq.macro-eq2_3}) when starting from the initial state $r=0$ and $\eta_{init}=0.6$. As shown in Figs. \ref{fig:6}$\bf{c}$ and \ref{fig:7}$\bf{c}$, near or at the critical points, Algorithm \ref{alg1} converged to the solutions of the MEs (\ref{eq.macro-eq2_1})(\ref{eq.macro-eq2_2})(\ref{eq.macro-eq2_3}).

The properties for the source signals from the Gamma ($+$) and bilateral Gamma ($\pm$) distributions (see Supplementary Figs. 5 and 6) are similar to those in Fig. \ref{fig:6} and \ref{fig:7}.

\begin{figure}[t]
 \begin{center}
 \includegraphics[height=15cm]{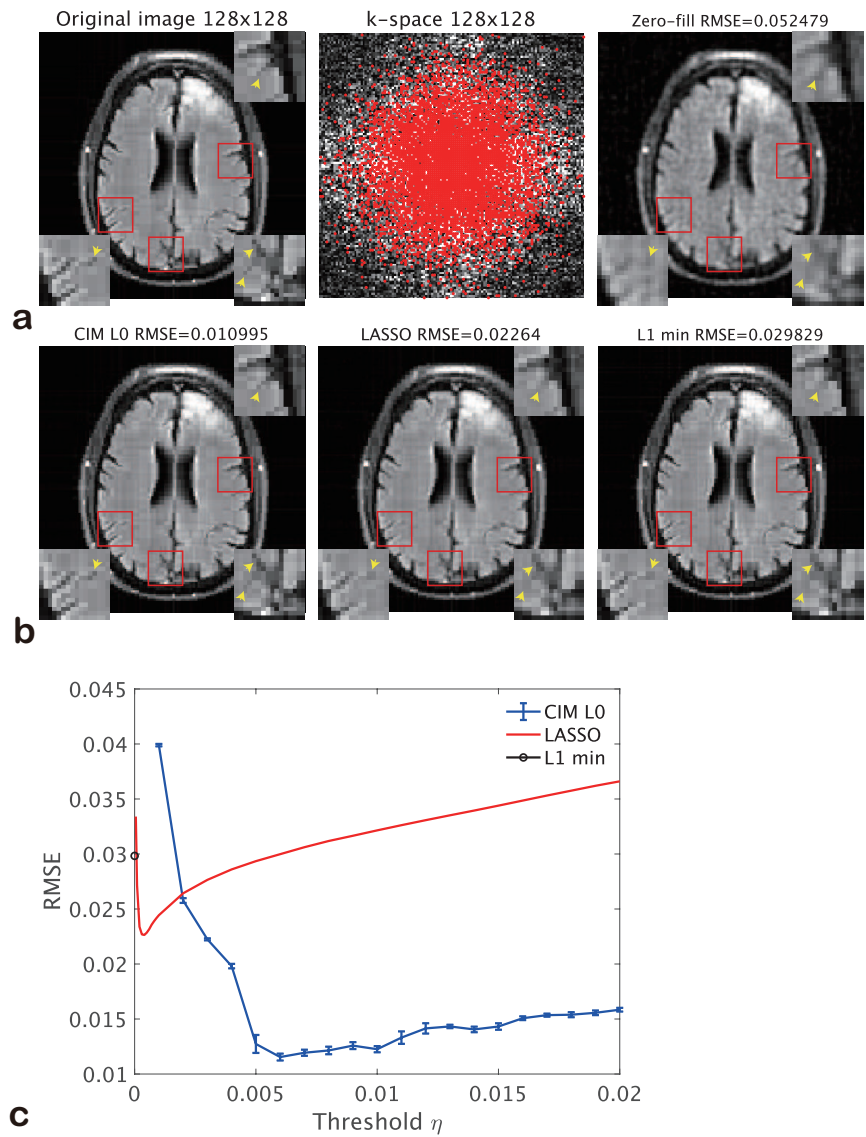}
 \end{center}
 \caption{Performance of CIM L0-RBCS and other methods on realistic data. $\bf{a}$ Left: Original image consisting of $128 \times 128$ pixels, which is spanned by Haar basis functions. The sparseness of the original image is $0.134$. Middle: k-space data ($128 \times 128$ pixels) obtained by performing a discrete Fourier transform on the original image. 30$\%$ of the k-space data were undersampled at random red points. Thus, the compression rate of the observation signal is $0.3$. Right: Zero-filling Fourier reconstruction from undersampled k-space data. $\bf{b}$ Reconstructed images with lowest errors and their RMSEs. Left: CIM L0-RBCS. $\eta_{init}=\eta_{end}= 0.004$. The initial state was given by LASSO. Middle: LASSO. $\eta=0.0004$. Right: L1 minimization-based CS. Inset figures: Enlarged regions labeled by red rectangles.    Yellow arrows point to pixel-level differences in the images.   $\bf{c}$ RMSEs as a function of the threshold $\eta$. Blue line with error bars: CIM L0-RBCS. Ten trials. Red line: LASSO. Circle: L1 minimization-based CS. For all methods, $\gamma=0.0001$.}
\label{fig:8}
\end{figure}

\subsection{Performance of CIM L0-RBCS on realistic data}
We evaluated the performance of CIM L0-RBCS and other methods on realistic data. We used MRI data obtained from the fastMRI datasets \cite{RN1000}. A Haar-wavelet transform (HWT) was applied to the data, and 86.6$\%$ of the HWT coefficients were set to zero to create a signal spanned by Haar basis functions with a sparseness of $0.134$ (left panel of Fig. \ref{fig:8}$\bf{a}$). The k-space data shown in the middle panel of Fig. \ref{fig:8}$\bf{a}$ was obtained by calculating the discrete Fourier transform (DFT) from the signal of the left panel of Fig. \ref{fig:8}$\bf{a}$, and 40$\%$ of the k-space data were undersampled at random red points in the middle panel of Fig. \ref{fig:8}$\bf{a}$ to create an observation signal with a compression rate of $0.4$. The right panel of Fig. \ref{fig:8}$\bf{a}$ shows an image with incoherent artifacts obtained by zero-filling Fourier reconstruction from the randomly undersampled k-space data.

To achieve higher reconstruction accuracy from the undersampled signal, we formulated the following implementable optimization problem on a CIM with L0 and L2 norms \cite{dedieu2020sampleefficient}:
\begin{eqnarray}
 x = \argmin_{x\in\mathbb{R}^N} \left(\frac{1}{2}\left\|y - S F x \right\|^2_2 + \frac{1}{2} \gamma \left\|\Delta_v x \right\|^2_2 + \frac{1}{2} \gamma \left\|\Delta_h x \right\|^2_2 + \lambda \left\|\Psi x \right\|_0 \right), \nonumber 
\end{eqnarray}
where $x$ is a source signal, $y$ is a k-space undersampling signal, $F$ is a DFT matrix, $S$ is an undersampling matrix, $\Psi$ is a HWT matrix, $\Delta_v$ and $\Delta_h$ are respectively the second-derivative matrices for the vertical and horizontal directions, and $\gamma$ and $\lambda$ are regularization parameters.    Under the variable transformation $r=\Psi x$, the mutual interaction matrix $J$ and the Zeeman term $h_z$ for CIM L0-RBCS are set as
\begin{eqnarray}
J &=& D \tilde{J} D, \ \ h_z = D S F \Psi^T y, \nonumber \\
& & \tilde{J}=\Psi F^T S^T S F \Psi^T + \gamma \Psi \Delta_v^T \Delta_v \Psi^T + \gamma \Psi \Delta_h^T \Delta_h \Psi^T, \nonumber \\
& & D = \left[ \begin{array}{ccc}
1/\sqrt{\tilde{J}_{11}} & & \Huge{0} \\
& \ddots & \\
\Huge{0} & & 1/\sqrt{\tilde{J}_{NN}} \\
\end{array} \right], \nonumber 
\end{eqnarray}
where $\tilde{J}_{ii}$ is a diagonal element of $\tilde{J}$ and $D$ is a diagonal matrix to normalize all diagonal elements of $\tilde{J}$ to $1$. Note that under the conversion described in Eq. (\ref{stationary_3}) and \ref{sec.maxwell}, all diagonal elements of the mutual interaction matrix $J$ need to be $1$. After the reconstruction with CIM L0-RBCS, $r'$, which is the output of the CDP, is transformed to the original scale signal $r$ with $r= D r'$.   

Furthermore, we evaluated the performance of LASSO minimizing $\frac{1}{2}\left\|y - S F x \right\|^2_2 + \frac{1}{2} \gamma \left\|\Delta_v x \right\|^2_2 +\frac{1}{2} \gamma \left\|\Delta_h x \right\|^2_2 + \lambda \left\|\Psi x \right\|_1$ and that of L1 minimization-based CS minimizing $\left\|\Psi x \right\|_1+ \gamma' \left\|\Delta_v x \right\|^2_2+\gamma' \left\|\Delta_h x \right\|^2_2$ s.t. $y = S F x$.

Figure \ref{fig:8}$\bf{b}$ shows images (and RMSEs) reconstructed from   Algorithm \ref{alg1} with $A_S^2=10^7$  (left panel of Fig. \ref{fig:8}$\bf{b}$), LASSO \cite{RN911} (middle panel of Fig. \ref{fig:8}$\bf{b}$), and L1-minimization-based CS implemented in CVX \cite{RN998,RN999} (right panel of Fig. \ref{fig:8}$\bf{b}$). As indicated in the images surrounded by the red circles in these panels, CIM L0-RBCS gave the most accurate reconstruction.

We evaluated the RMSEs of the three methods as a function of the threshold $\eta$. As shown in Fig. \ref{fig:8}$\bf{c}$, the blue line with error bars is the RMSE of CIM L0-RBCS obtained from ten trials, the red line is the RMSE of LASSO, and the circle is the RMSE of L1 minimization-based CS. There is an optimal value of $\eta$ to minimize the RMSEs of both CIM L0-RBCS and LASSO because of the trade-off between detecting small non-zero elements and eliminating incoherent artifacts by thresholding. The RMSE of CIM L0-RBCS was lower than those of the other methods in a wide range of $\eta$.

\begin{figure}[t]
 \begin{center}
 \includegraphics[height=10cm]{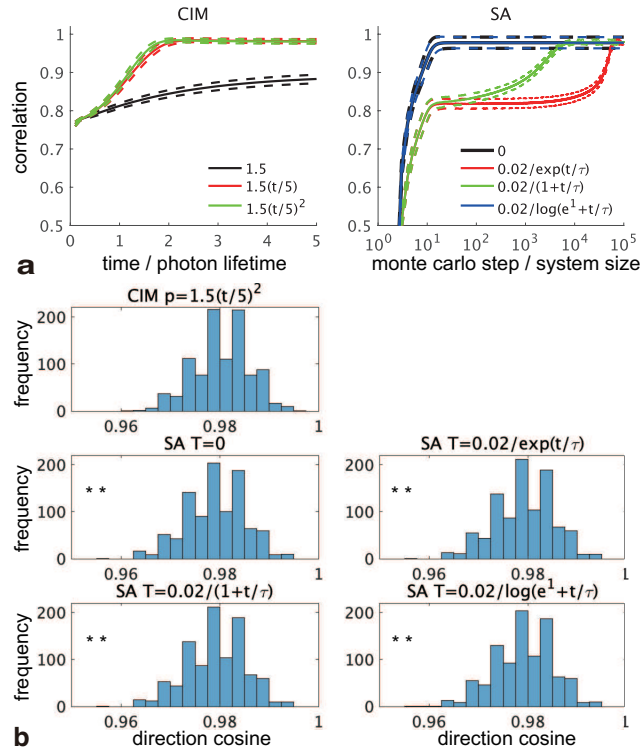}
 \end{center}
 \caption{  Comparison of CIM with SA in support vector estimation. {\bf a} Temporal profiles of support vector retrieval of CIM and SA. Threshold $\eta=0.05$ corresponding to L0-regularization parameter $\lambda = \eta^2/2=0.00125$. Left: Temporal change in direction cosine between the true support vector and one estimated with the CIM ($A_s^2=10^7$) under various rising schedule of pump rate. 1.5: constant (1.5). $1.5(t/5)$: linear rising. $1.5(t/5)^2$: square rising. In all cases, pump rates in the final state of $t = 5$ are 1.5. Right: Temporal change in direction cosine between the true support vector and estimated one with SA for various temperature lowering schedules. 0: constant (zero). $0.02/\exp(t/\tau)$: exponential cooling. $0.02/(1+t/\tau)$: inverse linear cooling. $0.02/\log(e^1+t/\tau)$: inverse log cooling. Except for the case of zero temperature, the initial temperature at $t=0$ is 0.02, and each $\tau$ is set so that the final temperature at $t=10^5$ is 0.00002. Both graphs show the mean (solid line) and standard deviation (dashed line) of 1000 samples. {\bf b} Distribution of direction cosines of the final state in the CIM and SA. Upper left: Histogram of final direction cosines at $t = 5$ of 1000 samples obtained from the CIM under square-rising pump-rate schedule. Others: Histograms of final direction cosines at $t=10^5$ of 1000 samples obtained from SA at zero temperature, exponential cooling, inverse linear cooling and inverse log cooling schedules. The two-sample one-sided Kolmogorov-Smirnov test suggests that the cumulative histograms for SA are significantly larger than that of the CIM, and thus, the histogram of the final direction cosines of the CIM is significantly biased toward right side compared with those of SA (Alternative hypothesis). ** in the graphs means P-value $< 0.01$. In both {\bf a} and {\bf b}, the observation matrix and the source signal and the true support vector were synthesized according to the precondition for applying statistical mechanics (Section \ref{precondition}). $N=500$, $\alpha=a=0.6$, $\beta=0$. Gaussian signal ($\pm$). In both the CIM and SA, $r_i$ was given the source signal $x_i$.}
\label{fig:9}
\end{figure}

\begin{algorithm}[t]
\caption{  Monte Carlo algorithm for support estimation}
\begin{algorithmic}[1]
\REQUIRE $M$-by-$N$ observation matrix: $A$, $M$-dimensional observation signal: $y$, $N$-dimensional signal vector: $r$
 \ENSURE $N$-dimensional support vector: $\sigma$
 \STATE Initialize $\sigma=0$
 \FOR{t=0 to $10^5 N$}
 \STATE Update the temperature $T$\\
 \STATE Randomly choose the spin index $i$ between $1$ and $N$ uniformly\\
 \STATE Calculate the acceptance ratio
\begin{eqnarray}
L&=&\exp\left(-\frac{1}{T} \left(\mathcal{H}(\sigma_1,\cdots,1-\sigma_i,\cdots,\sigma_N)-\mathcal{H}(\sigma_1,\cdots,\sigma_i,\cdots,\sigma_N) \right)\right) \nonumber \\ 
&=&\exp\left(\frac{1}{2T} \left(1-2\sigma_i\right)\left( - r_i^2 \sum_{\mu=1}^M {A_i^\mu}^2 + 2 r_i h_i - 2 \lambda \right) \right).\nonumber
\end{eqnarray}
\STATE Generate a uniform random number $U$ in the interval $[0, 1)$\\
\STATE Update the spin variable: 
\begin{eqnarray}
 \sigma_i = \left\{ \begin{array}{lc}
1-\sigma_i & {\rm if} L > U\\
\sigma_i & {\rm if} L \leq U
\end{array}\right.. \nonumber
\end{eqnarray}
 \ENDFOR
 \RETURN $\sigma$
\end{algorithmic}
\label{alg2}
\end{algorithm}

\subsection{Comparison of CIM with simulated annealing}
\label{sec.SA}
  
To demonstrate the efficacy of the CIM, we compared its ability to estimate support vectors with that of simulated annealing (SA).

Algorithm \ref{alg2} is the Monte Carlo algorithm we used for the support vector estimation in L0-RBCS. Here, $h_i$ is the local field given in Eq. (\ref{local_orgin}), and $2\lambda$ is equal to $\eta^2$, as described in Section \ref{sub2.1} and \ref{sec.maxwell}. To improve the estimation accuracy, the threshold $\eta$ corresponding to $\lambda$ needs to be set to a small finite value, as shown in Sections \ref{sub.Perturbation} and \ref{sec.phase_dia}. However, when $\lambda$ is small, the Monte Carlo algorithm cannot retrieve the support vector until the temperature is low enough to allow the L0-regularization term to work as a sparse bias. For the L0-regularization parameter of $\lambda=0.00125$ corresponding to $\eta=0.05$, we selected the initial and the final temperature at time $t=0$ and $t=10^5$ (Monte Carlo step/N) (see Supplementary material). Except for the zero-temperature case, we set the initial and final temperature to $0.02$ and $0.00002$, respectively.

In the experiment, 1000 samples of the observation matrix and source signal and true support vector were randomly synthesized according to the precondition for applying statistical mechanics (Section \ref{precondition}) under the Gaussian signal condition ($\pm$), $N=500$, $\alpha=a=0.6$ and $\beta=0$. By sharing of the same random seed, the same samples of the observation matrix and source signal and support vector could be used in different conditions of the CIM and SA. $r_i$ was given the source signal $x_i$. To measure the retrieval quality, we used the direction cosine between the true support vector $[\xi_1,\cdots,\xi_N]$ and the estimated one $[\sigma_1,\cdots,\sigma_N]$, which is defined as $\sum_{i=1}^N \xi_i \sigma_i / \sqrt{\sum_{i=1}^N \xi_i \sum_{i=1}^N \sigma_i}$. The direction cosine is 1 if the CIM (SA) perfectly retrieves the support vector. 

First, we evaluated the temporal profiles of the support vector retrievals of the CIM and SA under various pump-rate and temperature schedules. The left graph in Fig. \ref{fig:9}{\bf a} shows the temporal change in the direction cosine between the true support vector and the one estimated with the CIM ($A_s^2=10^7$, $\eta=0.05$) for constant, linear rising, and square rising schedules of the pump rate. In all cases, the pump rates in the final state are $1.5$. Each of the colored solid and dashed lines indicates the mean and standard deviation of 1000 samples. In the case of the constant pump rate, the direction cosine did not converge to 1 until $t = 5$ (time/photon lifetime), whereas in the cases of the linear and square rising schedules, it converged to about 1 around $t = 2$. On the other hand, the colored solid and dashed lines of the right graph for SA ($\lambda=0.00125$) show that the direction cosine converged to around 1 by $t=10^5$(Monte Carlo step / N) for all of the zero temperature, exponential, inverse linear, and inverse log cooling schedules. Note that the profile of the direction cosine of the inverse log cooling schedule is almost the same as that of the zero temperature case, because the temperature of the inverse log cooling schedule rapidly approaches the final temperature under the condition of the final time of $t=10^5$ (Monte Carlo step / N). Furthermore, the standard deviation of the direction cosine in all these cases was larger than those of the CIM. 

Next, we compared the distribution of the direction cosines of the final state in the CIM with those of SA under various cooling schedules. The upper left graph in Fig. \ref{fig:9}{\bf b} shows the histogram of the final direction cosines of 1000 samples obtained from the CIM for the square rising schedule of the pump rate, and the other graphs in Fig. \ref{fig:9}{\bf b} show histograms of the final direction cosines of 1000 samples obtained from SA for the zero temperature, exponential cooling, inverse linear cooling, and inverse log cooling schedules. These graphs suggest that the proportion of the direction cosines close to 1 in the 1000 samples of the CIM is higher than those of SA. The two-sample one-sided Kolmogorov-Smirnov test suggests that the histogram of the final direction cosines of the CIM is significantly biased toward the right side compared with all of those of SA (P-value $< 0.01$). Table \ref{table:p-value} summarizes the P-values for various sparsenesses $a$ and compression ratios $\alpha$. As shown in Table \ref{table:p-value}, the P-values for exponential, inverse linear and inverse log cooling schedules are slightly larger than those of the zero temperature in some cases. Therefore, the histograms of the final direction cosines for these cooling schedules are slightly biased toward the right side compared with the zero temperature in some cases. However, the histograms of the CIM are biased toward the right side compared with those of these cooling schedules; in particular the bias of the CIM is significant under conditions close to $\alpha=a$ (P-value$<0.05$).

The above results thus demonstrate that the CIM outperformed SA at support vector estimation. 
 
\begin{table}[h]
 \caption{ List of P-values of the two-sample one-sided Kolmogorov-Smirnov test for checking whether the histogram of the final direction cosines of the CIM is significantly biased toward the right side compared to those of SA (Alternative hypothesis). CIM ($p=1.5(t/5)^2$) vs. SA ($T=0$), CIM ($p=1.5(t/5)^2$) vs. SA ($T=0.02/\exp(t/\tau)$), CIM ($p=1.5(t/5)^2$) vs. SA ($T=0.02/(1+t/\tau)$) and CIM ($p=1.5(t/5)^2$) vs. SA ($T=0.02/\log(e^1+t/\tau)$). Each element of the table is a P-value for a certain sparseness $a$ and compression ratio $\alpha$.}
 \label{table:p-value}
 \centering

 \begin{tabular}{ccccccccc}
 \hline
 & \multicolumn{2}{c}{$T=0$} & \multicolumn{2}{c}{$T=0.02/\exp(t/\tau)$} &\multicolumn{2}{c}{$T=0.02/(1+t/\tau)$} & \multicolumn{2}{c}{$T=0.02/\log(e^1+t/\tau)$}\\
 \hline
 & $\alpha=0.4$ & $\alpha=0.6$ & $\alpha=0.4$ & $\alpha=0.6$ & $\alpha=0.4$ & $\alpha=0.6$ &  $\alpha=0.4$ & $\alpha=0.6$\\ 
 \hline\hline
 $a=0.1$ & 0.7206 & 0.6130 & 0.7466 & 0.5857 & 0.7466 & 0.5857 & 0.7206 & 0.6130 \\
 $a=0.2$ & 0.3325 & 0.4025 & 0.1983 & 0.4526 & 0.1983 & 0.4526 & 0.3325 & 0.4272 \\
 $a=0.3$ & 0.0798 & 0.0973 & 0.0469 & 0.1177 & 0.0420 & 0.1177 & 0.0973 & 0.0882 \\
 $a=0.4$ & 0.0000 & 0.0333 & 0.0007 & 0.0374 & 0.0003 & 0.0420 & 0.0000 & 0.0296 \\
 $a=0.5$ & 0.0000 & 0.0061 & 0.0000 & 0.0053 & 0.0000 & 0.0053 & 0.0000 & 0.0053 \\
 $a=0.6$ & 0.0000 & 0.0000 & 0.0000 & 0.0002 & 0.0000 & 0.0002 & 0.0000 & 0.0001 \\
 \hline 
 \end{tabular}
\end{table}

\section{Discussion}
\subsection{Summary and Conclusion}
We proposed a quantum-classical hybrid system that performs CIM and CDP steps alternately to optimize $r$ and $\sigma$. To evaluate the performance of CIM L0-RBCS, we introduced W-SDE as a model for a system consisting of $N$ OPOs and a measurement-feedback circuit. We obtained the MEs for CIM L0-RBCS from the W-SDE (\ref{eq.target-model}) and simultaneous equations (\ref{eq.simultaneous-equations1}).

As shown in Figs. \ref{fig:3}, \ref{fig:6}$\bf{c}$, and \ref{fig:7}$\bf{c}$ and Supplementary Figs. 1, 5$\bf{c}$ and 6$\bf{c}$, the theoretical results obtained from the MEs were consistent with the numerical results of Algorithm \ref{alg1} regardless of whether observation noise existed in the observed signal $y$. In particular, the theoretical results in the limit $A_s^2\to\infty$ were in good agreement with those of Algorithm \ref{alg1} with $A_s^2=10^7$. Because $A_s^2=10^7$ is on the same order as $A_s^2$ in the experimental CIMs \cite{RN899,RN813}, we expect that the MEs (\ref{eq.macro-eq2_1})(\ref{eq.macro-eq2_2})(\ref{eq.macro-eq2_3}) can be used to evaluate real experimental CIMs.

In the case of no observation noise, we theoretically showed that the performance of CIM L0-RBCS in principle approaches the threshold of L0-minimization-based CS \cite{Kabashima_2009,RN901} at high pump rates (see Fig. \ref{fig:4}$\bf{a}$ and Supplementary Fig. 3$\bf{b}$). From a mathematical perspective, the threshold $a=\alpha$ is the condition when the rank of a matrix composed of the column vectors of an observation matrix corresponding to the non-zero elements of the source signal is full. Thus, it is impossible for any system to go beyond this line mathematically. As described above, because the theoretical results in the limit $A_s^2\to\infty$ are in good agreement with those of Algorithm \ref{alg1} with $A_s^2=10^7$, we expect that the theoretical performance limit of real experimental CIMs will be close to this ideal limit.

In the case of observation noise, we theoretically showed that the RMSEs of CIM L0-RBCS are lower than those of LASSO for almost all conditions in which LASSO has an error less than $0.2$ and thus that CIM L0-RBCS exceeds LASSO's estimation accuracy under the optimal threshold for each method (see Figs. \ref{fig:6}$\bf{a}$, \ref{fig:6}$\bf{b}$, \ref{fig:7}$\bf{a}$ and \ref{fig:7}$\bf{b}$ and Supplementary Figs. 5$\bf{a}$, 5$\bf{b}$, 6$\bf{a}$ and 6$\bf{b}$).

However, there is a problem regarding the basin of attraction. As numerically demonstrated in Fig. \ref{fig:5} and Supplementary Figs. 4, when there is no observation noise, Algorithm \ref{alg1} cannot reach the theoretical performance limit if it starts from the practical initial condition $r=0$. However, even in such a situation, Algorithm \ref{alg1} exceeds LASSO's estimation accuracy until the lower bound of the critical points of CIM L0-RBCS (Fig. \ref{fig:5} and Supplementary Fig. 4). On the other hand, when there is observation noise, under the practical initial condition $r=0$, Algorithm \ref{alg1} gets very close to or achieves the theoretical performance limit of the ME (see Figs. \ref{fig:6}$\bf{c}$ and \ref{fig:7}$\bf{c}$ and Supplementary Figs. 5$\bf{c}$ and 6$\bf{c}$).

Finally, we confirmed using realistic data that CIM L0-RBCS gave the most accurate reconstruction compared with LASSO and L1-minimization-based CS (Fig. \ref{fig:8}).

Therefore, we can conclude that the performance of CIM L0-RBCS in principle approaches the theoretical limit of L0-minimization-based CS at high pump rates, exceeds that of LASSO, and moreover in practical situations exceeds LASSO's estimation accuracy.

A detailed interpretation and discussion of these results is given below.

\subsection{Effectiveness of CIM in support estimation}
  
As shown in Fig. \ref{fig:9}, the CIM outperformed SA in support estimation. In particular, as shown in Table \ref{table:p-value}, its superiority was significant under conditions close to the critical-point line $\alpha=a$. Close to the critical-point line $\alpha=a$, the energy landscape becomes more complicated. Therefore, this result indicates that the CIM can retrieve a support vector more efficiently than SA, especially in situations where the energy landscape is complicated near the critical point.

To improve the estimation accuracy of L0-RBCS, $\eta$ corresponding to the L0-regularization parameter $\lambda$ needs to be set to a small finite value. However, when $\lambda$ is small, the Monte Carlo algorithm cannot retrieve the support vector until the temperature is low enough to allow the L0-regularization term to work as a sparse bias. As described in Section \ref{sec.SA}, there is no remarkable improvement in SA comparable to the CIM. This result suggests that SA may not work well in such a situation where thermal fluctuations must be small like this. On the other hand, the CIM searches for the ground state on the basis of the minimum gain principle \cite{RN846,RN939}, which is different from thermal relaxation. Therefore, the results in Fig. \ref{fig:9} and Table \ref{table:p-value} demonstrate that the CIM is effective at solving a combinatorial optimization problem in such a situation where the thermal fluctuation must be small.
 
\subsection{Correctness of assumptions}
To derive the MEs (\ref{eq.macro-eq1_1})(\ref{eq.macro-eq1_2})(\ref{eq.macro-eq1_3}), we derived an approximate value for $\left<\tilde{X}(\tilde{h},t)\right>_{\rm SDE}$ of each OPO pulse by replacing the state variables in the second-order coefficient of the power of the quantum noise with average values of the state variables (see Eq. (\ref{eq.average-model})). As shown in Figs. \ref{fig:3}$\bf{b}$, \ref{fig:6}$\bf{c}$, and \ref{fig:7}$\bf{c}$, the ME derived under this approximation has good accuracy at the values of $A_s^2$ used in the actual CIM equipment. However, as shown in Fig. \ref{fig:3}$\bf{a}$, some solutions of the ME did not match the numerical solutions of Algorithm \ref{alg1} for smaller values of $A_s^2$. Thus, this approximation is possible if the mutual injection field is much larger than the noise in the steady state where the c-amplitude has grown.

\subsection{Basin of attraction and its dependency on the threshold}
To make the basin of attraction of Algorithm \ref{alg1} wider, we heuristically introduced a linear threshold attenuation in which the threshold $\eta$ linearly decreases as the alternating minimization proceeds. We confirmed that the basin of attraction widens as a result of lowering $\eta$ from a higher initial threshold $\eta_{init}$ to a lower terminal threshold $\eta_{end}$ (see Fig. \ref{fig:5} and Supplementary Fig. 4).

According to the definition of the injection field for each OPO pulse in Eq. (\ref{eq.optical-injection-field}), the threshold $\eta$ acts as an external field to give a negative bias for the OPO pulses to take the down state. By initially giving a large negative external field, almost all of the OPO pulses take the $\pi$-phase state, and thus, almost all of the $\{H(X_j)\}_{j=1,\cdots,N}$ take zero in the initial stage of the alternating minimization process. In the initial stage, the system can easily reach the ground state under a strong negative bias because the phase space, which consists of a small number of up-state OPO pulses, is simple. Then, through the alternating minimization process, the system tracks gradual changes in the ground state due to incremental increases in the number of up-state OPO pulses by gradually sweeping out a negative external field. Finally, the system achieves the ground state at the terminal threshold $\eta_{end}$. 

However, as demonstrated in Fig. \ref{fig:5}$\bf{b}$, when there is no observation noise, the system fails to converge to the near-zero-RMSE solutions beyond the lower bound line of the critical points. We suspect that there might be many quasi steady states beyond the lower bound line, as in the spin-glass phase \cite{RN941}; thus, the system might become trapped in one of the quasi steady states.

On the other hand, when there is observation noise, as demonstrated in Figs. \ref{fig:6}$\bf{c}$ and \ref{fig:7}$\bf{c}$, the system converges to near-zero-RMSE solutions even nearby the critical point when it starts from the practical initial condition $r=0$. It was suggested that the symmetries of the system allow for the creation of quasi steady states \cite{RN942}. We conjecture that observation noise could break the symmetries for quasi steady states.

\subsection{Plan to improve CIM L0-RBCS}
  
In this study, we used a W-SDE corresponding to the macroscopic model of MFB-CIM proposed by \cite{RN819,haribara2017performance}. On the other hand, there is a microscopic model, called the Gaussian approximation model, that provides a better approximation of the measurement process \cite{inui2020noise}. Moreover, we should mention that more general quantum models of the MFB-CIM without the Gaussian approximation have been derived for both discrete time models \cite{RN914} and continuous time models \cite{RN1474}. In future work, we will need to use these more general quantum models to evaluate the performance of CIM-L0-RBCS.

Furthermore, we will need to construct a full quantum system in which both the support estimation and the signal estimation are implemented on the CIM. We expect that due to the minimum gain principle, the full quantum system simulated with more general quantum models could overcome the quasi-steady-state problem discussed above.

\ack

This work is supported by the Japan Science and Technology Agency through its ImPACT program and NTT Research, Inc. All authors acknowledge the support of the NSF CIM Expedition award (CCF-1918549).

\appendix

\section{Derivation of Eqs. (5)-(8)}
\label{sec.maxwell}
  
The gradient of the Hamiltonian $\mathcal{H}$ with respect to each of $\sigma$ and $r$ is simply derived as
\begin{eqnarray}
-\frac{\partial \mathcal{H}}{\partial r_i}&=&-r_i\sigma_i^2\sum_{\mu=1}^M {A_i^\mu}^2 + \sigma_i h_i, \label{eq.a1}\\
-\frac{\partial \mathcal{H}}{\partial \sigma_i}&=&-r_i^2\sigma_i\sum_{\mu=1}^M {A_i^\mu}^2 + r_i h_i -\lambda, \label{eq.a2}\\
& &h_i =-\sum_{j=1(\neq i)}^N \sum_{\mu=1}^M A_i^\mu A_j^\mu \sigma_j r_j + \sum_{\mu=1}^M A_i^\mu y^\mu. \label{eq.a3}
\end{eqnarray}
Here, $h_i$ is the same as the local field defined in Eq. (\ref{local_orgin}). Since $r_i^2\geq 0$, $\sum_{\mu=1}^M {A_i^\mu}^2>0$, and $h_i$ in Eq. (\ref{eq.a3}) does not include $\sigma_i$, Eq. (\ref{stationary_1}) can be obtained from Eq. (\ref{eq.a2}) at $-\frac{\partial \mathcal{H}}{\partial \sigma_i}=0$ as follows.
\begin{eqnarray}
H\left(r_i^2\sigma_i\sum_{\mu=1}^M {A_i^\mu}^2\right)=\sigma_i = H\left(r_i h_i - \lambda\right). \nonumber
\end{eqnarray}
Here, $H(X)$ is the Heaviside step function taking $0$ for $X\leq 0$ or $+1$ for $X>0$. If $r_i=0$, the sign of $r_i h_i-\lambda$ is negative and $r_i^2 \sigma_i \sum_{\mu=1}^M {A_i^\mu}^2 = 0$. Thus, $\sigma_i$ consistently becomes zero if $r_i=0$. $\sigma_i$ takes either $0$ or $1$ depending on the sign of $r_i h_i - \lambda$.

\begin{figure}[t]
 \begin{center}
 \includegraphics[height=8cm]{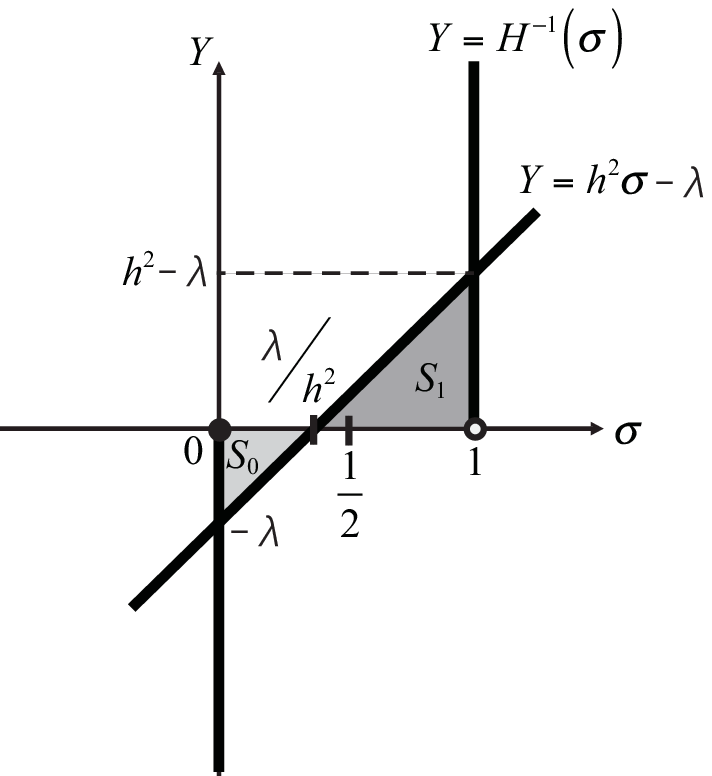}
 \end{center}
 \caption{  Selection of a stable fixed point according to the Maxwell rule}
\label{fig:APP1}
\end{figure}

Furthermore, since $\sigma_i^2=\sigma_i$, the following equation can be obtained from Eq. (\ref{eq.a1}) at $-\frac{\partial \mathcal{H}}{\partial r_i}=0$.
\begin{eqnarray}
r_i\sigma_i \sum_{\mu=1}^M {A_i^\mu}^2 = \sigma_i h_i. \label{eq.a4}
\end{eqnarray}
Note that $r_i$ is indefinite in Eq. (\ref{eq.a4}) when $\sigma_i=0$. Because $r_i \sigma_i=0$ holds if $\sigma_i=0$, $r_i$ can be safely set to zero when $\sigma_i=0$. To satisfy $r_i=0$ if $\sigma_i=0$, we modify Eq. (\ref{eq.a4}) to Eq. (\ref{stationary_2}):
\begin{eqnarray}
r_i \sum_{\mu=1}^M {A_i^\mu}^2 = \sigma_i h_i. \nonumber
\end{eqnarray}
In this study, we assume that $\sum_{\mu=1}^M {A_i^\mu}^2=1$ is satisfied. This assumption does not lose any generality because it is possible to normalize the observation matrix $A$ to satisfy $\sum_{\mu=1}^M {A_i^\mu}^2=1$ for any case. Under this assumption, the following equation is obtained from Eq. (\ref{stationary_2}).
\begin{eqnarray}
r_i = \sigma_i h_i. \label{eq.a6}
\end{eqnarray}
Before eliminating $r_i$ with the following manipulation, one should notice that $r_i$ is a solution in the steady-state with respect to $r_i$ satisfying $-\frac{\partial \mathcal{H}}{\partial r_i}=0$. $h_i$ in Eq. (\ref{eq.a3}) does not include $\sigma_i$. Thus, $r_i$ is uniquely determined by $\sigma_i$ and $h_i$. Then, substituting Eq. (\ref{eq.a6}) into Eq. (\ref{stationary_1}), we obtain 
\begin{eqnarray}
\sigma_i = H\left(\sigma_i h_i^2 - \lambda\right). \label{eq.a7}
\end{eqnarray}
Equation (\ref{eq.a7}) is a self-consistent equation to determine the value of $\sigma_i$. Figure \ref{fig:APP1} shows a schematic Maxwell rule to solve the self-consistent equation (\ref{eq.a7}) for $\sigma_i$. As shown, there are two stable fixed points $(0,-\lambda)$ and $(1,h_i^2-\lambda)$ corresponding to the two crossing points of the functions $Y=h_i^2 \sigma_i-\lambda$ and $Y=H^{-1} (\sigma_i)$. The two areas $S_0$ and $S_1$ enclosed by $Y=h_i^2 \sigma_i-\lambda$ and $Y=H^{-1} (\sigma_i)$ correspond to the depth of microscopic energy at two stable fixed points $(\sigma_i,Y)=(0,-\lambda)$ and $(1,h_i^2 - \lambda)$. According to the Maxwell rule, we select the stable fixed point with the largest enclosed area. Which of $S_0$ and $S_1$ is larger is determined by whether $\lambda /h_i^2$ is larger or smaller than $1/2$. 

If the source signal is signed ($\chi=\pm$), a stationary point of Eq. (\ref{eq.a7}) is determined by the following equation.
 \begin{eqnarray}
 \sigma_i = \left\{ \begin{array}{lc}
1 & h_i>\sqrt{2\lambda} \; {\rm or}\; h_i<-\sqrt{2\lambda}\\
0 & {\rm otherwise}
\end{array}\right.. \label{eq.a8}
\end{eqnarray}
Note that if the source signal is signed ($\chi=\pm$), $\sigma_i=1$ holds for both the positive side ($h_i > \sqrt{2\lambda}$) and the negative side ($h_i<-\sqrt{2\lambda}$). On the other hand, if the source signal is non-negative ($\chi=+$), $\sigma_i=1$ must hold for only the positive side ($h_i > \sqrt{2\lambda}$) to keep $r_i$ non-negative. In this case, a stationary point of Eq. (\ref{eq.a7}) is determined by
\begin{eqnarray}
 \sigma_i = \left\{ \begin{array}{lc}
1 & h_i>\sqrt{2\lambda} \\
0 & {\rm otherwise}
\end{array}\right.. \label{eq.a9}
\end{eqnarray}
Eq. (\ref{stationary_3}) allows us to write a unified equation for Eqs. (\ref{eq.a8}) and (\ref{eq.a9}):
\begin{eqnarray}
 \sigma_i &=& H\left(F_\chi(h_i) - \sqrt{2 \lambda}\right), \nonumber \\
 & &F_\chi(h) = \left\{ \begin{array}{l}
h \;\;\; ( \chi = + ) \\
|h| \;\; ( \chi = \pm )
\end{array}\right.. \nonumber
\end{eqnarray}
Finally, we confirm that the Hamiltonian $\mathcal{H}$ decreases at each iteration of a sequential update rule based on Eq. (\ref{stationary_3}). The change in $\mathcal{H}$ due to the $i$-th Potts spin flipping $\sigma_i$ to $\sigma_i'$ is expressed by the following equation with substitution of Eq. (\ref{eq.a6}).
\begin{eqnarray}
& &\mathcal{H}(\sigma_1,\cdots,\sigma_i',\cdots,\sigma_N)-\mathcal{H}(\sigma_1,\cdots,\sigma_i,\cdots,\sigma_N) \nonumber \\
& &=-\frac{1}{2}(\sigma_i' - \sigma)(h_i^2 - 2 \lambda).\label{eq.a10}
\end{eqnarray}
Substituting $\sigma_i'=H\left(F_\chi(h_i) - \sqrt{2 \lambda}\right)$ into (\ref{eq.a10}) yields
\begin{eqnarray}
& &\mathcal{H}(\sigma_1,\cdots,\sigma_i',\cdots,\sigma_N)-\mathcal{H}(\sigma_1,\cdots,\sigma_i,\cdots,\sigma_N)\nonumber \\
& &=-\frac{1}{2}\left(H\left(F_\chi(h_i) - \sqrt{2 \lambda}\right)- \sigma\right)(h_i^2 - 2 \lambda).\label{eq.a11}
\end{eqnarray}
The case of $\chi=\pm$\\ If $\sigma_i=0$ and either $h_i>\sqrt{2\lambda}$ or $h_i<-\sqrt{2\lambda}$, $\mathcal{H}(\sigma_1,\cdots,\sigma_i',\cdots,\sigma_N)-\mathcal{H}(\sigma_1,\cdots,\sigma_i,\cdots,\sigma_N) =-1/2 (h_i^2-2\lambda)<0$. If $\sigma_i=1$ and $-\sqrt{2\lambda}\leq h_i \leq \sqrt{2\lambda}$, $\mathcal{H}(\sigma_1,\cdots,\sigma_i',\cdots,\sigma_N)-\mathcal{H}(\sigma_1,\cdots,\sigma_i,\cdots,\sigma_N)=1/2 (h_i^2-2\lambda) \leq 0$. If $\sigma_i'=\sigma_i$, $\mathcal{H}(\sigma_1,\cdots,\sigma_i',\cdots,\sigma_N)-\mathcal{H}(\sigma_1,\cdots,\sigma_i,\cdots,\sigma_N)=0$. \\
The case of $\chi=+$\\
If $\sigma_i=0$ and $h_i>\sqrt{2\lambda}$, $\mathcal{H}(\sigma_1,\cdots,\sigma_i',\cdots,\sigma_N)-\mathcal{H}(\sigma_1,\cdots,\sigma_i,\cdots,\sigma_N) =-1/2 (h_i^2-2\lambda)<0$. If $\sigma_i=1$ and $-\sqrt{2\lambda}\leq h_i \leq \sqrt{2\lambda}$, $\mathcal{H}(\sigma_1,\cdots,\sigma_i',\cdots,\sigma_N)-\mathcal{H}(\sigma_1,\cdots,\sigma_i,\cdots,\sigma_N)=1/2 (h_i^2-2\lambda) \leq 0$. If $\sigma_i'=\sigma_i$, $\mathcal{H}(\sigma_1,\cdots,\sigma_i',\cdots,\sigma_N)-\mathcal{H}(\sigma_1,\cdots,\sigma_i,\cdots,\sigma_N)=0$. The growth condition for $\mathcal{H}$: If $\sigma_i=1$ and $h_i<-\sqrt{2\lambda}$, $\mathcal{H}(\sigma_1,\cdots,\sigma_i',\cdots,\sigma_N)-\mathcal{H}(\sigma_1,\cdots,\sigma_i,\cdots,\sigma_N)=1/2 (h_i^2-2\lambda)>0$. Note that if $\sigma_i=1$, $h_i=r_i\geq 0$ holds because $r_i=\sigma_i h_i$ is satisfied and $r_i$ is non-negative. Thus, the growth condition for $\mathcal{H}$ cannot exist.

In conclusion, the Hamiltonian $\mathcal{H}$ decreases monotonically at each iteration of the sequential update rule for both $\chi=\pm$ and $\chi=+$.
 
\section{Derivation of W-SDE for CIM}
\label{sec.W-SDE}
As shown in Fig. \ref{fig:1}, the pump pulses are injected into the main ring cavity through a second harmonic generation (SHG) crystal. A periodically poled lithium niobate (PPLN) waveguide is a highly efficient nonlinear medium for optical parametric oscillation. Suppose that the amplitude of the pump field injected into the main cavity is $\epsilon$ and the parametric coupling constant of the PPLN waveguide between the signal field and the pump field is $\kappa$. Then, the pumping Hamiltonian is $\hat{\cal H}_1=i\hbar\epsilon (\hat{a}_p^\dagger -\hat{a}_p)$ and the parametric interaction Hamiltonian is $\hat{\cal H}_2=i\hbar\kappa/2 (\hat{a}_s^{\dagger 2}\hat{a}_p - \hat{a}_p^\dagger \hat{a}_s^2)$. Here, $\hat{a}_p$ and $\hat{a}_s$ are the annihilation operators for the intra-cavity pump and signal fields. If the round-trip time of the ring cavity is correctly adjusted to $N$ times the pump pulse interval, $N$ independent and identical OPO pulses are simultaneously generated inside the cavity. The photon annihilation and creation operators for the $j$-th OPO signal pulse are denoted by $\hat{a}_j$ and $\hat{a}_j^{\dagger}$. The intra-cavity pump field and signal field have loss rates $\gamma_p$ and $\gamma_s$, respectively. If $\gamma_p\gg\gamma_s$, the pump field can be eliminated by invoking the slaving principle: the following master equation of the density operator for a solitary $j$-th OPO signal pulse is obtained by adiabatic elimination of the pump mode \cite{RN904,RN876},
\begin{eqnarray}
 \frac{\partial\hat{\rho}_{OPO}}{\partial t} &=& -i\hbar \frac{S}{2} \sum_{j=1}^N [\hat{a}_j^{\dagger 2} - \hat{a}_j^2, \hat{\rho}_{OPO}] \nonumber \\
& & + \gamma_s \sum_{j=1}^N \left(2\hat{a}_j \hat{\rho}_{OPO} \hat{a}_j^{\dagger} - \hat{a}_j^{\dagger}\hat{a}_j\hat{\rho}_{OPO} - \hat{\rho}_{OPO} \hat{a}_j^{\dagger}\hat{a}_j\right)\nonumber\\
& &+\frac{B}{2} \sum_{j=1}^N\left(2\hat{a}_j^2\hat{\rho}_{OPO} \hat{a}_j^{\dagger 2} - \hat{a}_j^{\dagger 2}\hat{a}_j^2 \hat{\rho}_{OPO} - \hat{\rho}_{OPO} \hat{a}_j^{\dagger 2}\hat{a}_j^2 \right),\label{eq.master-equation}
\end{eqnarray}
where $S=\epsilon\kappa/\gamma_p$ and $B=\kappa^2/(2\gamma_p)$ are the linear parametric gain coefficient and two photon absorption (or back conversion) rate, respectively. $[\hat{x},\hat{y}]$ denotes the bosonic commutator.

Next, let us examine the measurement-feedback circuit shown in Fig. \ref{fig:1}. The circuit is connected to the main cavity by extraction and injection couplers with reflection coefficients $R_{ex}=j_{ex} \Delta t$ and $R_{in}=j_{in} \Delta t$, where $j_{ex}$ and $j_{in}$ are coarse-grained out-coupling and in-coupling constants and $\Delta t$ is the cavity round trip time. When $B/\gamma_s<<1$ and vacuum fluctuations are incident on the open ports of the extraction and injection couplers, the measurement-feedback circuit can be described with a Gaussian quantum model \cite{RN908,inui2020noise}. The master equation consists of a linear loss term, measurement-induced state reduction term, and coherent feedback signal injection term (see Eqs. (12)(13)(14) in ref. \cite{inui2020noise}).

The Fokker-Planck equation is derived using the Wigner $W(\alpha)$ representation of the density operator $\hat{\rho}$ in the master equations, and we arrive at the following truncated Wigner stochastic differential equation (W-SDE) by applying Ito's rule \cite{RN1477,inui2020noise},
\begin{eqnarray}
 \frac{d\alpha_i}{dt}&=&-(\gamma_s + j) \alpha_i + S\alpha_i^* - B|\alpha_i|^2\alpha_i \nonumber \\
 & &+ j_{in} f_i^{sig} + \sqrt{\frac{\gamma_s}{2} + \frac{j}{2} + B|\alpha_i|^2} \upsilon_i,\ (i=1,\cdots,N)\label{W-SDE1}
\end{eqnarray}
where $j=j_{ex}+j_{in}$, $\alpha_i$ is the complex Wigner amplitude, and $\upsilon_i$ is the c-number noise amplitude satisfying $\left<\upsilon_i (t)\right>=0$, $\left<\upsilon_i^*(t) \upsilon_j(t')\right>=2\delta_{ij}\delta(t-t')$.

Then, by introducing a saturation parameter $A_s=\sqrt{2\gamma_p (\gamma_s+j)/\kappa^2}$ and applying the following scale transformation: $\alpha_i/A_s =c_i+is_i$, $t(\gamma_s+j)=t$, $p=S/(\gamma_s+j)$ and $K j_{in}/A_s(\gamma_s+j)=\tilde{K}$, we obtain Eq. (\ref{eq.target-model}). 

\section{Mean-field behavior of OPO pulses and CDP}
\label{sub.ave}
We approximately calculate the conditional expectations of $\tilde{X}(\tilde{h},t)$, $\tilde{G}(\tilde{h},t)$ and $\tilde{G}(\tilde{h},t)^2$ given the pure local field $\tilde{h}$, which are denoted by $\left<\tilde{X}(\tilde{h},t)\right>_{\rm SDE}$, $\left<\tilde{G}(\tilde{h},t)\right>_{\rm SDE}$ and $\left<\tilde{G}(\tilde{h},t)^2\right>_{\rm SDE}$ \cite{RN872}.

Under the premise that the local field can be separated into the pure local field and the ORT (Eq. (\ref{eq.local-field-ORT})) by SCSNA \cite{RN849,RN848,RN916}, substituting Eq. (\ref{eq.local-field-ORT}) into Eq. (\ref{eq.simultaneous-equations1}) and because $\frac{1}{M}\sum_{\mu=1}^M \left(A_i^\mu\right)^2 = 1$, we can write $r_i$ as
\begin{eqnarray}
r_i = \left\{ \begin{array}{ll}
0 & c_i\le 0 \\
\frac{\tilde{h}_i}{1-\Gamma} &c_i>0
\end{array}\right..\label{eq.r_value1}
\end{eqnarray}
Furthermore, substituting Eqs. (\ref{eq.local-field-ORT}) and (\ref{eq.r_value1}) into the W-SDE (\ref{eq.target-model}) gives:
\begin{eqnarray}
 \frac{dc_i}{dt} &=& (-1+p-c_i^2-s_i^2)c_i + \frac{1}{A_s}\sqrt{c_i^2+s_i^2+1/2} W_{i,1} \nonumber\\
 & &+ \tilde{K}\left(F_\chi\left(\tilde{h}_i + \tilde{h}_i\frac{\Gamma H(c_i)}{1-\Gamma}\right)-\eta\right), \nonumber\\
\frac{ds_i}{dt} &=& (-1-p-c_i^2-s_i^2 )s_i+ \frac{1}{A_s}\sqrt{c_i^2+s_i^2+1/2} W_{i,2}.\ (i=1,\cdots,N)\label{eq.target-model2}
\end{eqnarray}
Equation (\ref{eq.target-model2}) of the $i$-th OPO pulse only depends on the pure local fields $\tilde{h}_i$, which are statistically independent of each other in the steady state. The W-SDE (\ref{eq.target-model2}) can be regarded as describing $N$ independent one-body OPO pulses in the steady state. Thus, it is not necessary to solve the W-SDE (\ref{eq.target-model2}) simultaneously.

Since the steady-state solution of Eq. (\ref{eq.target-model2}) depends only on the value of the pure local field, the site index $i$ in Eq. (\ref{eq.target-model2}) can be deleted. It is difficult to solve Eq. (\ref{eq.target-model2}) analytically even after the $N$-body system has been reduced to a one-body system. To obtain a mathematically tractable form, we replace the state variables in the second-order coefficient of the Kramers-Moyal expansion \cite{RN1477} (representing the power of the quantum noise) with the average values of these state variables \cite{RN872}:
\begin{eqnarray}
 \frac{dc}{dt} &=& (-1+p-c^2-s^2) c + \frac{1}{A_s} \sqrt{\left<c^2\right>+\left<s^2\right>+1/2} W_{1} \nonumber \\
 & &+ \tilde{K}\left(F_\chi\left(\tilde{h}+\tilde{h}\frac{\Gamma H(c)}{1-\Gamma}\right)-\eta\right), \nonumber \\
 \frac{ds}{dt} &=& (-1-p-c^2-s^2 ) s + \frac{1}{A_s} \sqrt{\left<c^2\right> + \left<s^2\right> + 1/2} W_{2}. \label{eq.average-model}
\end{eqnarray}
From Eq. (\ref{eq.average-model}), we can derive the following equations to determine the approximate value of $\left<\tilde{X}(\tilde{h},t)\right>_{\rm SDE}$ for a single OPO pulse \cite{RN872}:
\begin{eqnarray}
 & &\left<\tilde{X}(\tilde{h},t)\right>_{\rm SDE} = \int_{-\infty}^{+\infty} dc \int_{-\infty}^{+\infty} ds H(c) f(c,s|\tilde{h}),\nonumber \\
 & &f(c,s|\tilde{h})\propto\exp\left(\frac{2A_s^2\left(c\tilde{K}\left(F_\chi\left(\tilde{h}+\tilde{h}\frac{\Gamma H(c)}{1-\Gamma}\right)-\eta\right)-V(c,s)\right)}{\Xi_{c}(\tilde{h})+\Xi_{s}(\tilde{h})+0.5}\right),\nonumber \\
& &V(c,s)=\frac{1}{2}(1-p)c^2+\frac{1}{2}(1+p)s^2+\frac{1}{2}c^2s^2+\frac{1}{4}c^4+\frac{1}{4}s^4,\nonumber
\end{eqnarray}
where $V(c,s)$ is the potential appearing in the CIM-ferromagnetic and the CIM-finite loading Hopfield models \cite{RN872}. $\Xi_{c}$ and $\Xi_{s}$ are parameters for calculating $\left<\tilde{X}(\tilde{h},t)\right>_{\rm SDE}$, which satisfy
\begin{eqnarray}
\Xi_{c}(\tilde{h})&=&\int_{-\infty}^{+\infty}dc\int_{-\infty}^{+\infty}ds c^2 f(c,s|\tilde{h}), \nonumber \\
\Xi_{s}(\tilde{h})&=&\int_{-\infty}^{+\infty}dc\int_{-\infty}^{+\infty}ds s^2 f(c,s|\tilde{h}). \nonumber
\end{eqnarray}
$\Xi_{c}$ and $\Xi_{s}$ are equal to $\left<c^2\right>$ and $\left<s^2\right>$, and by giving $\tilde{h}$ and $\Gamma$, they can be self-consistently determined from the above equation.

Similarly, from Eq. (\ref{eq.r_value1}), $\left<\tilde{G}(\tilde{h},t)\right>_{\rm SDE}$ and $\left<\tilde{G}(\tilde{h},t)^2\right>_{\rm SDE}$ can be obtained as follows:
\begin{eqnarray}
 \left<\tilde{G}(\tilde{h},t)\right>_{\rm SDE} &=& \frac{\tilde{h}}{1-\Gamma}\left<\tilde{X}(\tilde{h},t)\right>_{\rm SDE}, \nonumber \\
 \left<\tilde{G}(\tilde{h},t)^2\right>_{\rm SDE} &=& \frac{\tilde{h}^2}{(1-\Gamma)^2}\left<\tilde{X}(\tilde{h},t)\right>_{\rm SDE}. \nonumber
\end{eqnarray} 

\section{Details of SCSNA for the whole hybrid system}
\label{sub.scsna}
Under the precondition described in Section \ref{precondition}, we separate the local field into the pure local field and the ORT (Eq. (\ref{eq.local-field-ORT})) with SCSNA \cite{RN849,RN848,RN916,RN898,RN897}, and reduce the $N$-body system composed of $N$ mutually coupled OPO pulses to an effective one-body system. After that, we derive the ME for the whole hybrid system.

Let us start by introducing the following parameters.
\begin{eqnarray}
g^\mu=\frac{1}{N} \sum_{j=1}^N A_j^\mu (G(h_j,t)-\xi_j x_j)-\sqrt{\alpha/N} n^\mu,\label{eq.ass_parameter}
\end{eqnarray}
Below, we assume that $g^\mu=O(1/\sqrt{N})$ $(\mu=1,\cdots,M)$ is satisfied, because, under the precondition, the correlation between $A_j^\mu$ and $G(h_j,t)-\xi_j x_j$ is $O(1/\sqrt{N})$ for any $\mu$ if the reconstruction succeeds.

Substituting Eq. (\ref{eq.ass_parameter}) into Eq. (\ref{eq.local-field3}) gives
\begin{eqnarray}
h_i=-\frac{1}{\alpha} \sum_{\mu=1}^M A_i^\mu g^\mu + r_i H(c_i),\label{eq.local-field-cross}
\end{eqnarray}
where the first term is the cross-talk noise part, and the second term is introduced to subtract the direct self-coupling term from the local field $h_i$ because Eq. (\ref{eq.local-field3}) does not contain the direct self-coupling.

Next, we split the local field into a signal term, independent Gaussian noise, and the ORT. $g^\mu$, as defined in Eq. (\ref{eq.ass_parameter}), recursively contains $A_j^\mu g^\mu$ in $G(h_j,t)$, so it is a factor causing correlation between OPO pulses. Because $g^\mu=O(1/\sqrt{N})$, we perform the following expansion on Eq. (\ref{eq.ass_parameter}):
\begin{eqnarray}
g^\mu=\frac{1}{N} \sum_{j=1}^N A_j^\mu (G(h_j^{(\mu)},t)-\xi_j x_j) - \sqrt{\frac{\alpha}{N}} n^\mu- \frac{a}{\alpha} g^\mu U^{(\mu)},\label{eq.ass_parameter2}
\end{eqnarray}
where $h_i^{(\mu)}$ is the cavity field \cite{doi:10.1142/0271} and $U^{(\mu)}$ is a macroscopic parameter called the susceptibility, which are given by
\begin{eqnarray}
h_i^{(\mu)} &=& -\frac{1}{\alpha}\sum_{\nu=1(\neq\mu)}^M A_i^\nu g^\nu + r_i H(c_i), \\
U^{(\mu)} &=& \frac{1}{aN}\sum_{j=1}^N \frac{\partial G(h_j^{(\mu)},t)}{\partial h_j^{(\mu)}}, \ \ \label{eq.susp1}
\end{eqnarray}
The cavity field $h_j^{(\mu)}$ does not contain $A_j^{\mu} g^\mu$, so $G(h_j^{(\mu)},t)$ is uncorrelated with $A_j^{\mu}$ and $U^{(\mu)}$ is also uncorrelated with $A_j^{\mu}$. The terms that cause the correlation between the OPO pulses are extracted by performing a first-order Taylor expansion around $g^\mu=0$ and these extracted terms form the third term on the right side of Eq. (\ref{eq.ass_parameter2}).

From Eq. (\ref{eq.ass_parameter2}), we redefine $g^\mu$ on the basis of the cavity fields $h_i^{(\mu)}$ $(i=1,\cdots,N)$ as follows:
\begin{eqnarray}
g^\mu=\frac{\alpha}{\alpha+aU^{(\mu)}}\left(\frac{1}{N}\sum_{j=1}^N A_j^\mu (G(h_j^{(\mu)},t) - \xi_j x_j) - \sqrt{\frac{\alpha}{N}} n^\mu \right),\label{eq.ass_parameter3}
\end{eqnarray}
The terms causing the correlation between OPO pulses in Eq. (\ref{eq.ass_parameter}) are converted into the scale coefficient $\alpha/(\alpha+aU^{(\mu)})$.

Substituting Eq. (\ref{eq.ass_parameter3}) into the crosstalk noise in Eq. (\ref{eq.local-field-cross}), we split up the local field into three terms, as follows:
\begin{eqnarray}
 & &h_i=\left<\frac{\alpha x_i \xi_i}{\alpha +a U^{(\mu)}}\right>_\mu + Z_i + \left<\frac{a U^{(\mu)}}{\alpha+a U^{(\mu)}}\right>_\mu r_i H(c_i),\label{eq.local-field-ORT2}\\
 & &Z_i=-\frac{1}{N} \sum_{\mu=1}^M \sum_{j=1(\neq i)}^N \frac{A_i^\mu A_j^\mu (G(h_j^{(\mu)},t)-\xi_j x_j)}{\alpha+aU^{(\mu)}}+\sqrt{\frac{\alpha}{N}} \sum_{\mu=1}^M \frac{A_i^\mu n^\mu}{\alpha+aU^{(\mu)}},\label{eq.noise-ORT}
\end{eqnarray}
where the first term is the signal term, $Z_i$ is Gaussian random noise defined by Eq. (\ref{eq.noise-ORT}), and the third term is the self-coupling term. Here, $\left<\cdot\right>_\mu$ denotes $\left<x^{(\mu)}\right>_\mu = \frac{1}{M}\sum_{\mu=1}^M x^{(\mu)}$. These three terms are obtained under the conditions $\left<A_i^\mu\right>=0$ and $\left<A_i^\mu A_j^\nu\right>=\delta_{ij} \delta_{\mu\nu}$, and $G(h_i^{(\mu)},t)$ and $U^{(\mu)}$ are uncorrelated with $A_i^{\mu}$. Moreover, the third term is obtained under the approximation $\left<G(h_i^{(\mu)},t)\right>_{\mu} = r_i H(c_i)$. From the central limit theorem, $Z_i$ becomes Gaussian random noise in the thermodynamic limit. The average of $Z_i$ and the covariance between $Z_i$ and $Z_j$ are
\begin{eqnarray}
\left<Z_i\right> &=& 0, \nonumber \\
\left<Z_i Z_j\right> &=& \delta_{ij} \alpha^2 \left<\frac{\frac{a}{\alpha} (Q^{(\mu)}-2 R^{(\mu)}+\left<x^2\right>_x) + \beta^2}{(\alpha + aU^{(\mu)})^2}\right>_\mu, \nonumber
\end{eqnarray}
where $R^{(\mu)}$ and $Q^{(\mu)}$ are macroscopic parameters that are respectively called the overlap and the mean square magnetization and are given by
\begin{eqnarray}
& &R^{(\mu)}=\frac{1}{aN} \sum_{j=1}^N x_j \xi_j G(h_j^{(\mu)},t),\label{eq.overlap1}\\
& &Q^{(\mu)}=\frac{1}{aN} \sum_{j=1}^N G(h_j^{(\mu)},t)^2.\label{eq.square_mag1}
\end{eqnarray}

Because $Z_i$ is statistically independent of $Z_j$ when $i \neq j$, the first and second terms in Eq. (\ref{eq.local-field-ORT2}) are statistically independent of those of other sites. 
The third term is the difference between the self-coupling term in the crosstalk noise rescaled by $\alpha/(\alpha+aU^{(\mu)})$ and the original one (the second term of R.H.S in Eq. (\ref{eq.local-field-cross})), and it represents self-feedback via other OPO pulses. Therefore, the first and second terms are the pure local field and the third term is the ORT. By comparing Eqs. (\ref{eq.local-field-ORT}) and (\ref{eq.local-field-ORT2}), $\tilde{h}_i$ and $\Gamma$ are determined as follows:
\begin{eqnarray}
\tilde{h}_i = \left<\frac{\alpha x_i \xi_i}{\alpha +a U^{\mu}}\right>_\mu + Z_i,\ \Gamma = \left<\frac{a U^{\mu}}{\alpha+a U^{\mu}}\right>_\mu.\nonumber
\end{eqnarray}
As explained in \ref{sub.ave}, substituting Eq. (\ref{eq.local-field-ORT}) into the W-SDE (\ref{eq.target-model}) reduces the $N$-body system to an effective one-body system. The W-SDE (\ref{eq.target-model2}) can be regarded as $N$ independent equations. The $i$-th independent equation in the W-SDE (\ref{eq.target-model2}) implies that $H(c_i)$ is a stochastic variable depending on the pure local field $\tilde{h}_i$ and time $t$ in the steady state. Thus, $X(h_i,t)$ and $G(h_i,t)$ can be redefined as $\tilde{X}(\tilde{h}_i,t)$ and $\tilde{G}(\tilde{h}_i,t)$, as shown in Eq. (\ref{eq.transfer_for_PLF}):
\begin{eqnarray}
H(c_i) = \tilde{X}(\tilde{h}_i,t), \ \ r_i= \tilde{G}(\tilde{h}_i,t) = \frac{1}{1-\Gamma} \tilde{h}_i \tilde{X}(\tilde{h}_i,t). \nonumber
\end{eqnarray}
Through the manipulations in Eqs. (\ref{eq.ass_parameter2}) and (\ref{eq.ass_parameter3}), the pure local field and the ORT are defined on the cavity field. In the thermodynamic limit ($N\rightarrow\infty$), the cavity field can be consistently replaced with the pure local field and the ORT, and $G(h^{(\mu)}_i,t)$ in Eqs. (\ref{eq.overlap1}) (\ref{eq.square_mag1}) (\ref{eq.susp1}) can be safely replaced with $\tilde{G}(\tilde{h}_i,t)$. As a result of this replacement, the cavity indexes $(\mu)$ of $R^{(\mu)}$, $Q^{(\mu)}$, and $U^{(\mu)}$ become negligible, and these macroscopic parameters are redefined with Eqs. (\ref{eq.overlap2}), (\ref{eq.square_mag2}) and (\ref{eq.susp2}):
\begin{eqnarray}
& &R=\frac{1}{aN} \sum_{j=1}^N x_j \xi_j \tilde{G}(\tilde{h}_i,t), \nonumber \\
& &Q=\frac{1}{aN} \sum_{j=1}^N \tilde{G}(\tilde{h}_i,t)^2, \nonumber \\
& &U=\frac{1}{aN} \sum_{j=1}^N \frac{\partial \tilde{G}(\tilde{h}_j,t)}{\partial \tilde{h}_j} \frac{\partial \tilde{h}_j}{\partial h_j},\nonumber
\end{eqnarray}
where $U$ expresses the average sensitivity of $\tilde{G}(\tilde{h}_i,t)$ to the bare local field $h_i$ using the chain rule because of the definition of $U^{(\mu)}$ in Eq. (\ref{eq.susp1}).

Because the pure local fields are independent of each other, the site averages in Eqs. (\ref{eq.overlap2})(\ref{eq.square_mag2})(\ref{eq.susp2}) can be replaced with the averages of $\left<\tilde{G}(\tilde{h},t)\right>_{\rm SDE}$ and $\left<\tilde{G}(\tilde{h},t)^2\right>_{\rm SDE}$ with respect to the Gaussian random noise $Z$ and the source signal $x \xi$. The replacement for $U$ can be achieved by integration by parts. Finally, we obtain the MEs (\ref{eq.macro-eq1_1})(\ref{eq.macro-eq1_2})(\ref{eq.macro-eq1_3}) for finite $A_s$ and the MEs (\ref{eq.macro-eq2_1})(\ref{eq.macro-eq2_2})(\ref{eq.macro-eq2_3}) for infinite $A_s$ in Section \ref{sec.outline}.

\newpage

\bibliographystyle{iopart-num}
\bibliography{L0-CIM}

\providecommand{\newblock}{}
\begin{thebibliography}{10}
\expandafter\ifx\csname url\endcsname\relax
  \def\url#1{{\tt #1}}\fi
\expandafter\ifx\csname urlprefix\endcsname\relax\def\urlprefix{URL }\fi
\providecommand{\eprint}[2][]{\url{#2}}

\bibitem{RN843}
Johnson M~W, Amin M~H, Gildert S, Lanting T, Hamze F, Dickson N, Harris R,
  Berkley A~J, Johansson J, Bunyk P, Chapple E~M, Enderud C, Hilton J~P, Karimi
  K, Ladizinsky E, Ladizinsky N, Oh T, Perminov I, Rich C, Thom M~C, Tolkacheva
  E, Truncik C~J, Uchaikin S, Wang J, Wilson B and Rose G 2011 {\em Nature\/}
  {\bf 473} 194--198

\bibitem{farhi2014quantum}
Farhi E, Goldstone J and Gutmann S 2014 A quantum approximate optimization
  algorithm \urlprefix\url{https://arxiv.org/abs/1411.4028}

\bibitem{PhysRevX.10.021067}
Zhou L, Wang S~T, Choi S, Pichler H and Lukin M~D 2020 {\em Phys. Rev. X\/}
  {\bf 10}(2) 021067
  \urlprefix\url{https://link.aps.org/doi/10.1103/PhysRevX.10.021067}

\bibitem{RN1589}
Goto H 2016 {\em Scientific Reports\/} {\bf 6} 21686
  \urlprefix\url{https://www.ncbi.nlm.nih.gov/pubmed/26899997}

\bibitem{RN1591}
Goto H 2019 {\em Journal of the Physical Society of Japan\/} {\bf 88} ISSN
  0031-9015 1347-4073

\bibitem{RN1594}
Goto H, Tatsumura K and Dixon A~R 2019 {\em Science Advances\/} {\bf 5}
  eaav2372
  \urlprefix\url{https://advances.sciencemag.org/content/advances/5/4/eaav2372.full.pdf}

\bibitem{RN1596}
Mahboob I, Okamoto H and Yamaguchi H 2016 {\em Science Advances\/} {\bf 2}
  e1600236
  \urlprefix\url{https://advances.sciencemag.org/content/advances/2/6/e1600236.full.pdf}

\bibitem{RN846}
Marandi A, Wang Z, Takata K, Byer R~L and Yamamoto Y 2014 {\em Nature
  Photonics\/} {\bf 8} 937--942

\bibitem{RN892}
Yamamoto Y, Aihara K, Leleu T, Kawarabayashi K, Kako S, Fejer M, Inoue K and
  Takesue H 2017 {\em npj Quantum Information\/} {\bf 3} 49 ISSN 2056-6387
  \urlprefix\url{https://www.nature.com/articles/s41534-017-0048-9.pdf}

\bibitem{RN899}
Inagaki T, Haribara Y, Igarashi K, Sonobe T, Tamate S, Honjo T, Marandi A,
  McMahon P~L, Umeki T, Enbutsu K, Tadanaga O, Takenouchi H, Aihara K,
  Kawarabayashi K~I, Inoue K, Utsunomiya S and Takesue H 2016 {\em Science\/}
  {\bf 354} 603--606

\bibitem{RN813}
McMahon P~L, Marandi A, Haribara Y, Hamerly R, Langrock C, Tamate S, Inagaki T,
  Takesue H, Utsunomiya S, Aihara K, Byer R~L, Fejer M~M, Mabuchi H and
  Yamamoto Y 2016 {\em Science\/} {\bf 354} 614--617 ISSN 0036-8075
  \urlprefix\url{https://science.sciencemag.org/content/354/6312/614}

\bibitem{RN1547}
Leleu T, Yamamoto Y, McMahon P~L and Aihara K 2019 {\em Phys. Rev. Lett.\/}
  {\bf 122}(4) 040607
  \urlprefix\url{https://link.aps.org/doi/10.1103/PhysRevLett.122.040607}

\bibitem{RN1548}
Kako S, Leleu T, Inui Y, Khoyratee F, Reifenstein S and Yamamoto Y 2020 {\em
  Advanced Quantum Technologies\/} {\bf 3} 2000045
  \urlprefix\url{https://onlinelibrary.wiley.com/doi/abs/10.1002/qute.202000045}

\bibitem{RN1598}
Sutton B, Camsari K~Y, Behin-Aein B and Datta S 2017 {\em Scientific reports\/}
  {\bf 7} 44370 \urlprefix\url{https://www.ncbi.nlm.nih.gov/pubmed/28295053}

\bibitem{RN1600}
Tait A~N, de~Lima T~F, Zhou E, Wu A~X, Nahmias M~A, Shastri B~J and Prucnal P~R
  2017 {\em Scientific reports\/} {\bf 7} 7430
  \urlprefix\url{https://www.ncbi.nlm.nih.gov/pubmed/28784997}

\bibitem{RN1602}
Yoshimura C, Yamaoka M, Hayashi M, Okuyama T, Aoki H, Kawarabayashi K and
  Mizuno H 2015 {\em Scientific reports\/} {\bf 5} 16213
  \urlprefix\url{https://www.ncbi.nlm.nih.gov/pubmed/26586362}

\bibitem{7350099}
{Yamaoka} M, {Yoshimura} C, {Hayashi} M, {Okuyama} T, {Aoki} H and {Mizuno} H
  2016 {\em IEEE Journal of Solid-State Circuits\/} {\bf 51} 303--309

\bibitem{8118124}
{Zhang} J, {Chen} S and {Wang} Y 2018 {\em IEEE Transactions on Computers\/}
  {\bf 67} 604--616

\bibitem{ChihiroYoshimura2017}
Yoshimura C, Hayashi M, Okuyama T and Yamaoka M 2017 {\em International Journal
  of Networking and Computing\/} {\bf 7} 154--172

\bibitem{RN1605}
Aramon M, Rosenberg G, Valiante E, Miyazawa T, Tamura H and Katzgraber H~G 2019
  {\em Frontiers in Physics\/} {\bf 7} 1--14 ISSN 2296-424X

\bibitem{RN1585}
Neukart F, Compostella G, Seidel C, von Dollen D, Yarkoni S and Parney B 2017
  {\em Frontiers in ICT\/} {\bf 4} 1--6 ISSN 2297-198X

\bibitem{RN1571}
O'Malley D, Vesselinov V~V, Alexandrov B~S and Alexandrov L~B 2018 {\em PLoS
  One\/} {\bf 13} e0206653
  \urlprefix\url{https://www.ncbi.nlm.nih.gov/pubmed/30532243}

\bibitem{RN1587}
Bando Y, Susa Y, Oshiyama H, Shibata N, Ohzeki M, G\'omez-Ruiz F~J, Lidar D~A,
  Suzuki S, del Campo A and Nishimori H 2020 {\em Phys. Rev. Research\/} {\bf
  2}(3) 033369
  \urlprefix\url{https://link.aps.org/doi/10.1103/PhysRevResearch.2.033369}

\bibitem{RN848}
Aonishi T, Kurata K and Okada M 1999 {\em Physical Review Letters\/} {\bf 82}
  2800--2803

\bibitem{RN900}
Tibshirani R 1996 {\em Journal of the Royal Statistical Society Series
  B-Methodological\/} {\bf 58} 267--288

\bibitem{RN927}
Claerbout J~F and Muir F 1973 {\em Geophysics\/} {\bf 38} 826--844

\bibitem{RN926}
Taylor H~L, Banks S~C and Mccoy J~F 1979 {\em Geophysics\/} {\bf 44} 39--52

\bibitem{RN928}
Chapman N~R and Barrodale I 1983 {\em Geophysical Journal of the Royal
  Astronomical Society\/} {\bf 72} 93--100

\bibitem{RN929}
Iinuma T, Hino R, Uchida N, Nakamura W, Kido M, Osada Y and Miura S 2016 {\em
  Nature Communications\/} {\bf 7} 13506

\bibitem{RN910}
Lustig M, Donoho D and Pauly J~M 2007 {\em Magnetic Resonance in Medicine\/}
  {\bf 58} 1182--1195

\bibitem{RN930}
Doneva M and Mertins A 2016 {\em Mri: Physics, Image Reconstruction, and
  Analysis\/} {\bf 49} 51--71

\bibitem{RN931}
Lu W, Atkinson I~C and Vaswani N 2016 {\em Mri: Physics, Image Reconstruction,
  and Analysis\/} {\bf 49} 27--49

\bibitem{RN932}
Yamamoto T, Fujimoto K, Okada T, Fushimi Y, Stalder A~F, Natsuaki Y, Schmidt M
  and Togashi K 2016 {\em Investigative Radiology\/} {\bf 51} 372--378

\bibitem{RN902}
Honma M, Akiyama K, Uemura M and Ikeda S 2014 {\em Publications of the
  Astronomical Society of Japan\/} {\bf 66} 95 (1--14)

\bibitem{RN933}
Ramprasad R, Batra R, Pilania G, Mannodi-Kanakkithodi A and Kim C 2017 {\em Npj
  Computational Materials\/} {\bf 3} 54

\bibitem{RN934}
Nakada G, Igarashi Y, Lmai H and Oaki Y 2019 {\em Advanced Theory and
  Simulations\/} {\bf 2} 1800180

\bibitem{RN920}
Fu W~J~J 1998 {\em Journal of Computational and Graphical Statistics\/} {\bf 7}
  397--416

\bibitem{RN921}
Efron B, Hastie T, Johnstone I and Tibshirani R 2004 {\em Annals of
  Statistics\/} {\bf 32} 407--451

\bibitem{RN925}
Friedman J, Hastie T, Hofling H and Tibshirani R 2007 {\em Annals of Applied
  Statistics\/} {\bf 1} 302--332

\bibitem{RN912}
Bioucas-Dias J~M and Figueiredo M~A~T 2007 {\em IEEE Transactions on Image
  Processing\/} {\bf 16} 2992--3004

\bibitem{RN911}
Beck A and Teboulle M 2009 {\em Siam Journal on Imaging Sciences\/} {\bf 2}
  183--202 ISSN 1936-4954

\bibitem{RN924}
Boyd S, Parikh N, Chu E, Peleato B and Eckstein J 2011 {\em Foundations and
  Trends in Machine Learning\/} {\bf 3} 1--122

\bibitem{RN909}
Louizos C, Welling M and Kingma D~P 2017 Learning sparse neural networks
  through $l_0$ regularization \urlprefix\url{https://arxiv.org/abs/1712.01312}

\bibitem{RN901}
Nakanishi-Ohno Y, Obuchi T, Okada M and Kabashima Y 2016 {\em Journal of
  Statistical Mechanics: Theory and Experiment\/} {\bf 2016} 063302

\bibitem{RN936}
Chen S~S, Donoho D~L and Saunders M~A 2001 {\em SIAM review\/} {\bf 43}
  129--159

\bibitem{RN937}
Chartrand R 2007 {\em IEEE Signal Processing Letters\/} {\bf 14} 707--710

\bibitem{RN938}
Tropp J~A and Gilbert A~C 2007 {\em IEEE Transactions on Information Theory\/}
  {\bf 53} 4655--4666

\bibitem{Benders:1962vi}
Benders J~F 1962 {\em Numerische Mathematik\/} {\bf 4} 238--252
  \urlprefix\url{https://doi.org/10.1007/BF01386316}

\bibitem{RN1576}
Choi V 2008 {\em Quantum Information Processing\/} {\bf 7} 193--209 ISSN
  1573-1332

\bibitem{RN1578}
Choi V 2010 {\em Quantum Information Processing\/} {\bf 10} 343--353 ISSN
  1570-0755 1573-1332

\bibitem{RN939}
Hamerly R, Inagaki T, McMahon P~L, Venturelli D, Marandi A, Onodera T, Ng E,
  Langrock C, Inaba K, Honjo T, Enbutsu K, Umeki T, Kasahara R, Utsunomiya S,
  Kako S, Kawarabayashi K, Byer R~L, Fejer M~M, Mabuchi H, Englund D, Rieffel
  E, Takesue H and Yamamoto Y 2019 {\em Science Advances\/} {\bf 5} eaau0823

\bibitem{RN940}
Sherrington D and Kirkpatrick S 1975 {\em Physical Review Letters\/} {\bf 35}
  1792--1796

\bibitem{RN849}
Shiino M and Fukai T 1992 {\em Journal of Physics a-Mathematical and General\/}
  {\bf 25} L375--L381

\bibitem{RN916}
Aonishi T, Kurata K and Okada M 2002 {\em Physical Review E\/} {\bf 65} 046223

\bibitem{RN872}
Aonishi T, Mimura K, Utsunomiya S, Okada M and Yamamoto Y 2017 {\em Journal of
  the Physical Society of Japan\/} {\bf 86} 104002

\bibitem{RN898}
Aonishi T, Okada M, Mimura K and Yamamoto Y 2018 {\em Journal of Applied
  Physics\/} {\bf 124} 152129

\bibitem{RN897}
Aonishi T, Mimura K, Okada M and Yamamoto Y 2018 {\em Journal of Applied
  Physics\/} {\bf 124} 233102

\bibitem{doi:10.1073/pnas.0502258102}
Donoho D~L and Tanner J 2005 {\em Proceedings of the National Academy of
  Sciences\/} {\bf 102} 9452--9457 (\textit{Preprint}
  \eprint{https://www.pnas.org/doi/pdf/10.1073/pnas.0502258102})
  \urlprefix\url{https://www.pnas.org/doi/abs/10.1073/pnas.0502258102}

\bibitem{Kabashima_2009}
Kabashima Y, Wadayama T and Tanaka T 2009 {\em Journal of Statistical
  Mechanics: Theory and Experiment\/} {\bf 2009} L09003
  \urlprefix\url{https://doi.org/10.1088/1742-5468/2009/09/l09003}

\bibitem{RN905}
Donoho D~L, Maleki A and Montanari A 2009 {\em Proceedings of the National
  Academy of Sciences\/} {\bf 106} 18914--18919

\bibitem{RN917}
Nishimori H 2001 {\em Statistical physics of spin glasses and information
  processing : an introduction\/} International series of monographs on physics
  (Oxford ; New York: Oxford University Press)

\bibitem{doi:10.1126/sciadv.abe7953}
Goto H, Endo K, Suzuki M, Sakai Y, Kanao T, Hamakawa Y, Hidaka R, Yamasaki M
  and Tatsumura K 2021 {\em Science Advances\/} {\bf 7} eabe7953
  (\textit{Preprint}
  \eprint{https://www.science.org/doi/pdf/10.1126/sciadv.abe7953})
  \urlprefix\url{https://www.science.org/doi/abs/10.1126/sciadv.abe7953}

\bibitem{RN881}
Abu-Rgheff M~A 2007 {\em Introduction to CDMA wireless communications\/} 1st ed
  (Amsterdam ; Boston ; London: Academic)

\bibitem{PhysRevLett.88.024102}
Aonishi T and Okada M 2001 {\em Phys. Rev. Lett.\/} {\bf 88}(2) 024102

\bibitem{RN875}
Yoshida M, Uezu T, Tanaka T and Okada M 2007 {\em Journal of the Physical
  Society of Japan\/} {\bf 76} 054003

\bibitem{RN1000}
Zbontar J, Knoll F, Sriram A, Murrell T, Huang Z, Muckley M~J, Defazio A, Stern
  R, Johnson P, Bruno M, Parente M, Geras K~J, Katsnelson J, Chandarana H,
  Zhang Z, Drozdzal M, Romero A, Rabbat M, Vincent P, Yakubova N, Pinkerton J,
  Wang D, Owens E, Zitnick C~L, Recht M~P, Sodickson D~K and Lui Y~W 2018
  fastmri: An open dataset and benchmarks for accelerated mri
  \urlprefix\url{https://arxiv.org/abs/1811.08839}

\bibitem{dedieu2020sampleefficient}
Dedieu A, L{\'a}zaro-Gredilla M and George D 2020 Sample-efficient l0-l2
  constrained structure learning of sparse ising models
  \urlprefix\url{https://arxiv.org/abs/2012.01744}

\bibitem{RN998}
Grant M and Boyd S 2008 Graph implementations for nonsmooth convex programs
  {\em Recent Advances in Learning and Control\/} Lecture Notes in Control and
  Information Sciences ed Blondel V, Boyd S and Kimura H (Springer-Verlag
  Limited) pp 95--110

\bibitem{RN999}
Grant M and Boyd S 2014 {CVX}: Matlab software for disciplined convex
  programming, version 2.1 http://cvxr.com/cvx

\bibitem{RN941}
Tanaka F and Edwards S~F 1980 {\em Journal of Physics F-Metal Physics\/} {\bf
  10} 2769--2778

\bibitem{RN942}
Crisanti A and Sompolinsky H 1988 {\em Physical Review A\/} {\bf 37} 4865--4874

\bibitem{RN819}
Haribara Y, Utsunomiya S and Yamamoto Y 2016 {\em A Coherent Ising Machine for
  MAX-CUT Problems: Performance Evaluation against Semidefinite Programming and
  Simulated Annealing\/} (Tokyo: Springer Japan) book section Chapter 12, pp
  251--262 Lecture Notes in Physics

\bibitem{haribara2017performance}
Haribara Y, Ishikawa H, Utsunomiya S, Aihara K and Yamamoto Y 2017 {\em Quantum
  Science and Technology\/} {\bf 2} 044002

\bibitem{inui2020noise}
Inui Y and Yamamoto Y 2020 Noise correlation and success probability in
  coherent ising machines \urlprefix\url{https://arxiv.org/abs/2009.10328}

\bibitem{RN914}
Yamamura A, Aihara K and Yamamoto Y 2017 {\em Phys. Rev. A\/} {\bf 96}(5)
  053834 \urlprefix\url{https://link.aps.org/doi/10.1103/PhysRevA.96.053834}

\bibitem{RN1474}
Shoji T, Aihara K and Yamamoto Y 2017 {\em Phys. Rev. A\/} {\bf 96}(5) 053833
  \urlprefix\url{https://link.aps.org/doi/10.1103/PhysRevA.96.053833}

\bibitem{RN904}
Kinsler P and Drummond P~D 1991 {\em Phys Rev A\/} {\bf 43} 6194--6208

\bibitem{RN876}
Maruo D, Utsunomiya S and Yamamoto Y 2016 {\em Physica Scripta\/} {\bf 91}
  083010

\bibitem{RN908}
Wiseman H~M and Milburn G~J 1993 {\em Phys Rev Lett\/} {\bf 70} 548--551

\bibitem{RN1477}
Risken H 1989 {\em The Fokker-Planck Equation Methods of Solution and
  Applications\/} second edition. ed Springer Series in Synergetics, (Berlin,
  Heidelberg: Springer Berlin Heidelberg,) ISBN 9783642615443 0172-7389 ;
  \urlprefix\url{http://dx.doi.org/10.1007/978-3-642-61544-3}

\bibitem{doi:10.1142/0271}
Mezard M, Parisi G and Virasoro M 1986 {\em Spin Glass Theory and Beyond\/}
  (WORLD SCIENTIFIC)

\end{thebibliography}

\end{document}